\documentclass[]{aa}  
\pdfoutput=1
\usepackage{graphicx}
\usepackage{txfonts}
\usepackage{natbib}
\usepackage{subcaption}
\usepackage{color}
\usepackage{threeparttable, tablefootnote}
\usepackage[LGRgreek]{mathastext}
\usepackage{chemformula}
\usepackage{gensymb}
\usepackage[utf8]{inputenc}
\usepackage[T1]{fontenc}
\usepackage{mhchem}

\begin{document} 

\title{Sulphur monoxide exposes a potential molecular disk wind from the planet-hosting disk around HD100546}
\author{Alice Booth \inst{1}, Catherine Walsh \inst{1}, Mihkel Kama \inst{2}, Ryan A. Loomis \inst{3}, Luke T. Maud \inst{4} \& Attila Juh\'asz \inst{3}} 
\institute{School of Physics and Astronomy, University of Leeds, Leeds LS2 9JT, UK \hfill \break \email{pyasb@leeds.ac.uk, C.Walsh1@leeds.ac.uk} 
\and Institute of Astronomy, Madingley Rd, Cambridge, CB3 0HA, UK
\and Harvard-Smithsonian Center for Astrophysics, 60 Garden Street, Cambridge, MA 02138, USA
\and Leiden Observatory, Leiden University, PO Box 9513, 2300 RA Leiden, The Netherlands}

\date{Received 12/06/2017; Accepted 12/12/2017}

\titlerunning{Detection of Sulphur Monoxide in the HD100546 Disk}
\authorrunning{Booth et al. 2017}

\abstract{
Sulphur-bearing volatiles are observed to be significantly depleted in interstellar and circumstellar regions. This 
missing sulphur is postulated to be mostly locked up in refractory form.
With ALMA we have detected sulphur monoxide (SO), a known shock tracer, in the HD 100546 protoplanetary disk.
Two rotational transitions: $J=7_{7}-6_{6}$ (301.286~GHz) and $J=7_8-6_7$ (304.078~GHz)
are detected in their respective integrated intensity maps. The 
stacking of these transitions results in a clear 5$\sigma$ detection in the stacked line profile.
The emission is compact but is spectrally resolved and the line profile has two components.
One component peaks at the source velocity and the other is blue-shifted by 5~km~s$^{-1}$. 
The kinematics and spatial distribution of the SO emission are not consistent with that expected from a purely Keplerian disk. 
We detect additional blue-shifted emission that we attribute to a disk wind.
The disk component was simulated using LIME and a physical disk structure.
The disk emission is asymmetric and best fit by a wedge of emission in the north east region of the disk
coincident with a `hot-spot' observed in the CO $J=3-2$ line. 
The favoured hypothesis is that a possible inner disk warp (seen in CO emission) directly exposes the north-east side of the disk
to heating by the central star, creating locally the conditions to launch a disk wind. Chemical 
models of a disk wind will help to elucidate why the wind is particularly highlighted in SO emission and 
whether a refractory source of sulphur is needed.
An alternative explanation is that the SO is tracing an accretion shock from a circumplanetary disk associated with the 
proposed protoplanet embedded in the disk at 50~au.
We also report a non-detection of SO in the protoplanetary disk around HD 97048.}


\maketitle

\section{Introduction}

Investigating the chemical structure and evolution of protoplanetary disks is important when studying
planet formation as the composition of a planet is determined by the content of its parent disk. 
The physical and chemical conditions in disks can be traced with observations of molecular line emission (see \citealt{Henning2013}, \citealt{2014prpl.conf..317D}, and references therein). 
These observations are key to understanding disk chemistry resulting from changes in disk structure
due to planet-disk interactions.

Transition disks were the first disks identified to possess signatures of clearing by unseen planets \citep{Strom1989}. Protoplanet 
candidates have been identified in the cavities of some transition disks e.g. LkCa 15 \citep{Sallum:2015ej} and 
HD 169142 \citep{Reggiani:2014dj}. 
These disks are expected to have a rich observable chemistry due to the disk midplane being directly exposed 
to far-UV radiation from the central star. Molecules that otherwise would be frozen out onto grains in the cold, 
shielded midplane of the disk may be detectable \citep{Cleeves2011}. 
High spatial resolution observations with ALMA allow investigation into the 
physical and chemical conditions associated with planet-forming regions of nearby disks.

The dominant volatiles (gas and ice) in protoplanetary disks are composed of hydrogen, oxygen, carbon, nitrogen and sulphur.
Sulphur is observed to be significantly depleted in circumstellar regions with 
detections of gas-phase S-bearing molecules only accounting for $\sim$0.1\% of the estimated cosmic abundance 
\citep[][]{Tieftrunk1994, Dutrey1997, Ruffle1999}. The missing sulphur is thought to reside in or on the
solid dust grains in the disk, the most abundant molecule being $\mathrm{H_{2}S}$ in cometary ices \citep{Bockel2000} and 
also in iron sulphides, another main component of primitive comets and meteorites \citep{Keller2002}.
The lack of agreement between observed abundances and chemical models points towards as yet unaccounted for grain-surface processes 
preventing desorption of $\mathrm{H_{2}S}$ and the formation of gas phase S-bearing species \citep{Dutrey2011}. 
This is supported by non-detections of anticipated molecules despite deep targeted searches in disks \citep{Martin2016}.

Despite the depletion issue, sulphur-bearing species are useful tracers of physical processes in interstellar and circumstellar 
material.
For example, SO is frequently detected as a tracer of shocked gas associated with the bipolar outflows from Class 0 and I protostars 
e.g., \citealt{Tafalla2010, Podio2015} and \citealt{Sakai2016}. Searches for SO have found that detections in outflows are ubiquitous 
but in protoplanetary disks infrequent \citep{Guilloteau2013}. The first detection of SO in a circumstellar disk was reported by \citet{Fuente2013}. They 
observed the $J=3_{4}-2_{3}$ (138.178~GHz) transition in the transitional disk AB Aur. This detection
was confirmed by \citet{Pacheco2015} with detections of the higher frequency $J=5_{4}-4_{3}$ (206.176~GHz) and $J=5_{6}-4_{5}$ (219.949~GHz)
transitions. Further observations and modelling suggest that the SO is distributed in a ring (145 to 384~au) and is depleted in the region of the 
disk's horse-shoe shaped dust trap \citep{Pacheco2016}. Most recently, \citet{Guilloteau2016} report the detection of the SO $J=6_{7}-5_{4}$ (251.826~GHz) and $J=6_{7}-5_{6}$ (261.978~GHz) transitions in four 
disks out of 30 that were observed with the IRAM 30-m telescope. The aforementioned studies show that SO has been detected
in both T Tauri and Herbig disks. ALMA observations have revealed ring components of SO emission in Class 0 protostars
\citep{Ohashi2014, Podio2015, Sakai2016} that have been interpreted as accretion shocks at the disk-envelope interface.

It is evident from existing observations that SO is a tracer of shocks in Class 0 and I protostars. 
However, SO is an elusive disk molecule 
requiring high sensitivity observations for its detection: this is now possible with ALMA.
We present the first detection of SO in the planet-hosting disk around HD 100546 from ALMA Cycle 0 observations
gaining insight into the molecular content and structure of transition disks. 
We also report a non-detection of SO in the protoplanetary disk around HD 97048.
The rest of this paper is structured as follows. Section 2 gives an overview 
of previous observations of HD 100546 and Section 3 describes our observations. 
The detected line emission and analysis are detailed in Sections 4 and 5. In Section 6 the results are discussed
and we list our conclusions and prospects for further work.   

\section{HD 100546}

HD 100546 is a $2.4~M_\odot$ Herbig Be star \citep{Ancker1998} at a distance of $109^{+~4}_{-~3}$~$pc$ \citep{Gaia1, Gaia2}
and host to a bright disk with a position angle of $146^{\circ}$ and an inclination of $44^{\circ}$  \citep[e.g.][]{Walsh2014}.
This transition disk is well observed because of its interesting structure and proximity.

Modelling of the SED and interferometric observations have shown that there is a cavity in the disk within 10~au of the central star  
and an inner dust disk ($\leq1~au$) located at around the sublimation temperature (1500~K) for silicate grains 
\citep{Bouwman2003, Benisty2010, Tatulli2011}. The inner disk was confirmed by 
the detection of the [OI] (6300\AA) line emission and CO ro-vibrational 
transitions that are coincident with the anticipated inner dust disk \citep{Acke2006, VanderPlas2009}.
There is an absence of gas observed in the dust cavity proposed to be due to gap clearing by a planet \citep{Brittain2009, Mulders2013}.

ALMA Band 7 observations show that the millimeter-sized dust grains are distributed in two rings: 
one 21~au wide centred at 26~au and one 75~au wide centred at 190~au \citep{Walsh2014}.
The best fit evolutionary model for the data is a scenario where there are two planets in the disk:
a $20~M_{J}$ planet at 10~au and $15~M_{J}$ planet at 68~au. 
Further modelling of the evolution of this system suggests that the inner planet
formed first, $\geq1~Myr$ into the disk's lifetime and that the second planet is 2~-~3~Myr younger and still
in the process of forming. This constraint is required in order to reproduce the contrast ratio in the observations between the inner and outer rings 
 \citep{Pinilla2015}. This is in agreement with observations of an embedded protoplanet 
and a possible companion in the cavity 
of the disk \citep[][]{Quanz2013, Mulders2013, Avenhaus2014, Brittain2014, Currie2014, Currie2015, Quanz2015, Benisty2016, Garufi2016}.

Various molecular and atomic lines have been detected with single-dish
telescopes. Although spatially unresolved, these observations suggest a warm disk atmosphere above the 
disk midplane with the highest abundances of the species coinciding with the outer edge of the cavity which is thought to be puffy
\citep{Strum2010, Panic2010,  Carmona2011,  Menard2011, Goto2012, Liskowsky2012, Fedele2013,  Bertelsen2014}.
The disk hosts a large molecular gas disk with the CO gas 
extending out to $\approx$~400~au \citep{2014ApJ...788L..34P, Walsh2014}.
In addition to this, there is evidence for thermal decoupling of gas and dust in the disk atmosphere 
and radial drift of millimeter-sized dust grains \citep[see][]{ Bruderer2012, Meeus2013, Walsh2014}.

\section{Observations}

The observations of HD 100546 and the four SO transitions analysed in this work are detailed in Table \ref{table:SOlines} (ALMA program 2011.0.00863.S).
The continuum and CO line emission from this data have already been analysed \citep[see][for full details]{Walsh2014} and 
this work uses the self-calibrated, phase-corrected and continuum subtracted measurements sets. 
Imaging of the data was done using Common Astronomy Software Application (CASA) version 4.6.0.
The individual lines were each imaged at the spectral resolution of the observations (0.24~km~s$^{-1}$, applying Hanning smoothing) 
and then at a coarser resolution of 1~km~s$^{-1}$. Only the $J=7_{7}-6_{6}$ and $J=7_8-6_7$ transitions were detected. 
These two SO lines were then stacked in the \textit{uv} plane to increase S/N
in the resulting channel maps (see Table 1).
This was done as follows. The central frequency of each spectral window was transformed to the central 
frequency of the SO line it contained using the CASA tool \texttt{regridspw}. 
These measurement sets were then concatenated using the CASA task \texttt{concat}. The line emission was
imaged at a velocity resolution of $1~km~s^{-1}$ using the CLEAN algorithm with natural weighting and a channel by channel mask guided by the 
spatial extent of the CO $J=3-2$ emission (3$\sigma$).
The $J=3_2-1_2$ transition has a low excitation
energy (21~K) and we expect the inner-most region of this disk to be warm ($>20~K$).
Hence, the detection of the higher energy transitions only is consistent with the expected temperature 
of emitting molecular gas.  Further, the lower energy transition is significantly weaker (see Table 1). 
 Including the $J=8_8-7_7$ transition in the stacking increased the noise level in the channel maps thus 
 degrading the S/N in the resulting images. 
Stacking wholly in the image plane gave the same results but with a slightly lower S/N than using the \texttt{regrid} plus \texttt{concat} method (also seen in the analysis of \citealt{Walsh2016}).

\begin{table*}[]
\centering
\begin{threeparttable}
\caption{ALMA band 7 observational parameters and sulphur monoxide transitions for HD 100546}
\label{table:SOlines}
\begin{tabular}{cccccc}
\hline\hline
\multicolumn{1}{l}{Date observed}                        			& \multicolumn{5}{c}{18th November 2012} \\
\multicolumn{1}{l}{Baselines}                            			& \multicolumn{5}{c}{21 - 375 m}     \\ 
\multicolumn{1}{l}{Weighting}                            			& \multicolumn{5}{c}{natural}      \\ \hline
\multicolumn{1}{l}{SO rotational transitions}            			& \multicolumn{1}{c}{$7_{7}-6_{6}$}                    & \multicolumn{1}{c}{$7_8-6_7$}                      		& \multicolumn{1}{c}{$8_8-7_7$}   							& \multicolumn{1}{c}{$3_2-1_2$}   								& \multicolumn{1}{c}{$7_{7}-6_{6}$ + $7_8-6_7$} \\
\multicolumn{1}{l}{Rest frequency (GHz)}                 			& \multicolumn{1}{c}{301.286}                          & \multicolumn{1}{c}{304.078}                        		& \multicolumn{1}{c}{344.311} 								& \multicolumn{1}{c}{345.704} 									&\multicolumn{1}{c}{-} \\
\multicolumn{1}{l}{Synthesised beam}                    			& \multicolumn{1}{c}{1\farcs1~$\times$~0\farcs6}       & \multicolumn{1}{c}{1\farcs1~$\times$~0\farcs6}             & \multicolumn{1}{c}{1\farcs0~$\times$~0\farcs5}            & \multicolumn{1}{c}{1\farcs0~$\times$~0\farcs5}            	& \multicolumn{1}{c}{1\farcs1~$\times$~0\farcs6} \\
\multicolumn{1}{l}{Beam P.A.}                       	 			& \multicolumn{1}{c}{24\degree}                        & \multicolumn{1}{c}{23$\degree$}                            & \multicolumn{1}{c}{40$\degree$}            				& \multicolumn{1}{c}{40$\degree$}            					& \multicolumn{1}{c}{24$\degree$} \\
\multicolumn{1}{l}{Spectral resolution (~km~s$^{-1}$)}              & \multicolumn{1}{c}{0.24}                             & \multicolumn{1}{c}{0.24}                                   & \multicolumn{1}{c}{0.21}            						& \multicolumn{1}{c}{0.21}           						    & \multicolumn{1}{c}{1.00} \\
\multicolumn{1}{l}{r.m.s noise (channel$^{-1}$ mJy beam$^{-1}$)}    & \multicolumn{1}{c}{10.9}                             & \multicolumn{1}{c}{9.9}                                    & \multicolumn{1}{c}{17.4}            						& \multicolumn{1}{c}{16.2}            							& \multicolumn{1}{c}{4.2} \\
\multicolumn{1}{l}{Peak emission (mJy beam$^{-1}$)}                 & \multicolumn{1}{c}{-}                                & \multicolumn{1}{c}{-}                                      & \multicolumn{1}{c}{-}            							& \multicolumn{1}{c}{-}            								& \multicolumn{1}{c}{24.7} \\ 
\multicolumn{1}{l}{E$_u$ (K)}                                       & \multicolumn{1}{c}{71.0}                             & \multicolumn{1}{c}{62.1}                                   & \multicolumn{1}{c}{87.5}            					    & \multicolumn{1}{c}{21.1}            						    & \multicolumn{1}{c}{-} \\ 
\multicolumn{1}{l}{Einstein A coefficient (s$^{-1}$)}               & \multicolumn{1}{c}{3.429e-04}                        & \multicolumn{1}{c}{3.609e-04}                              & \multicolumn{1}{c}{5.186e-04}            					& \multicolumn{1}{c}{1.390e-07}          						& \multicolumn{1}{c}{-} \\ \hline
\end{tabular}
\begin{tablenotes}\footnotesize
\item{The values for the line frequencies and Einstein A coefficients are from the Cologne Database for Molecular Spectroscopy \citep[CDMS;][]{2001A&A...370L..49M}
and the Leiden Atomic and Molecular Database \citep[LAMDA;][]{2005A&A...432..369S}.}
\end{tablenotes}
\end{threeparttable}
\end{table*}

\section{Results}
\subsection{Detected SO emission in the HD 100546 disk}

Figure 1 shows the integrated intensity maps of the two SO lines detected in the imaging. 
They encompass the significant ($>3\sigma$) on source emission detected in the $1~km~s^{-1}$ channel maps.
This is emission across 11 channels from $-7~km~s^{-1}$ to $5~km~s^{-1}$ with respect to the source velocity.
The source velocity of the emission, as inferred from the CO $J=3-2$ emission \citep{Walsh2014, walsh2017},
is 5.7~km~s$^{-1}$. The $J=7_{7}-6_{6}$ and $J=7_8-6_7$ transitions are detected with a peak S/N of 7 and
11 respectively in the integrated intensity. The r.m.s. noise was extracted from the region beyond the 3$\sigma$ contour of the integrated intensity.

Stacking these two transitions results in a convincing 
detection of SO in the channel maps (6$\sigma$) and line profile (5$\sigma$).
The line profiles of the individual transitions, at a channel width of 1~km~s$^{-1}$,
are shown in Appendix A. 
Figure 2 shows the channel maps of the stacked SO emission at a velocity resolution of 1~km~s$^{-1}$ and 
with respect to the source velocity.
The emission reaches a peak S/N of 6 with an r.m.s. noise of 4.2~mJy~beam$^{-1}$ channel$^{-1}$.
The noise in the channel maps was determined by taking the r.m.s. of
the line-free channels either side of the significant emission.
The CO emission (>~3$\sigma$) is also plotted to allow a comparison between these two molecules.
When comparing the SO and CO emission, 
the SO emission is significantly more compact than the CO emission and there is an excess blue-shifted
component of SO emission that is not spatially consistent with the blue-shifted disk emission traced in CO.
This is further highlighted by comparing the SO and CO line profiles shown in Appendix B.

\begin{figure*}
\includegraphics[width=0.5\hsize]{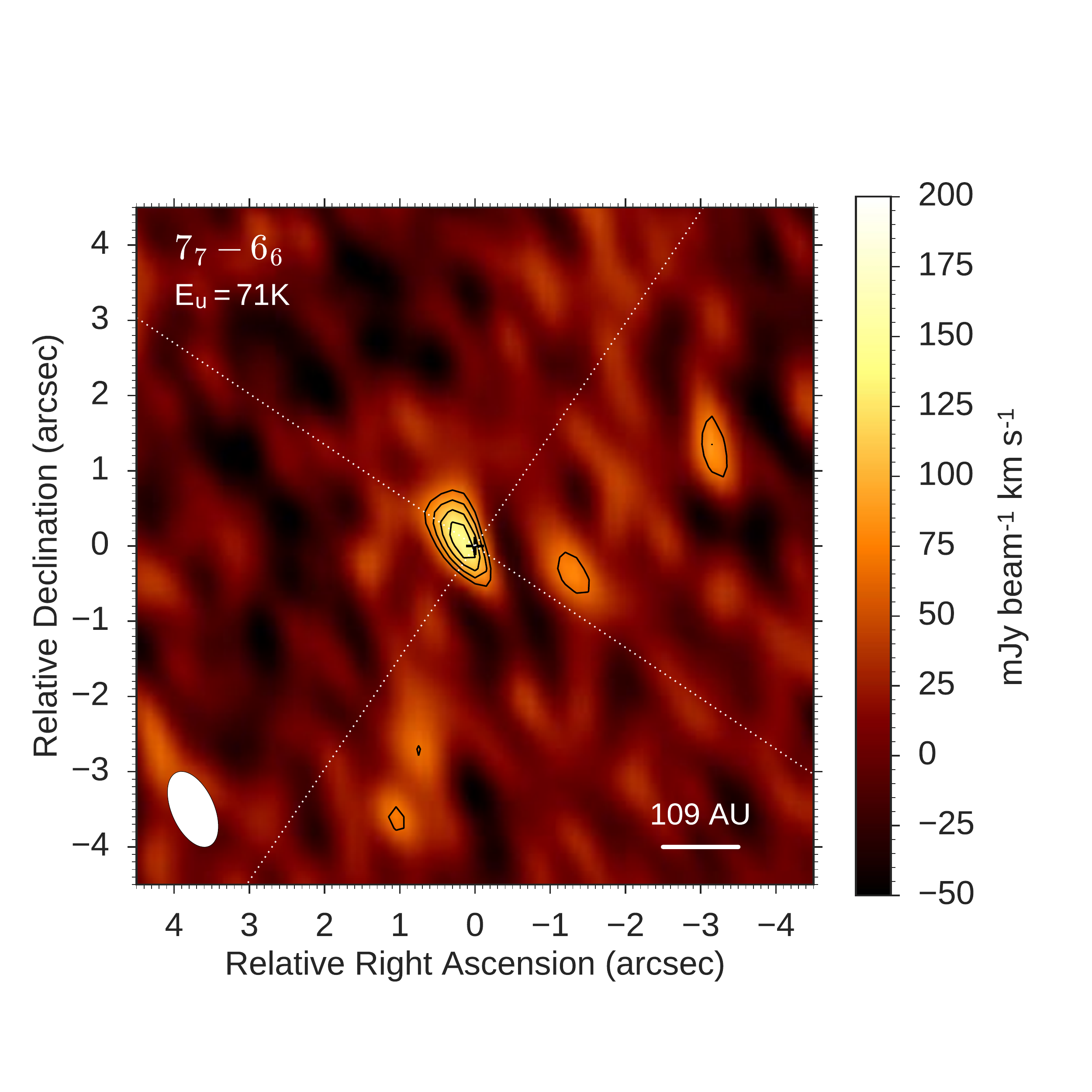}
\includegraphics[width=0.5\hsize]{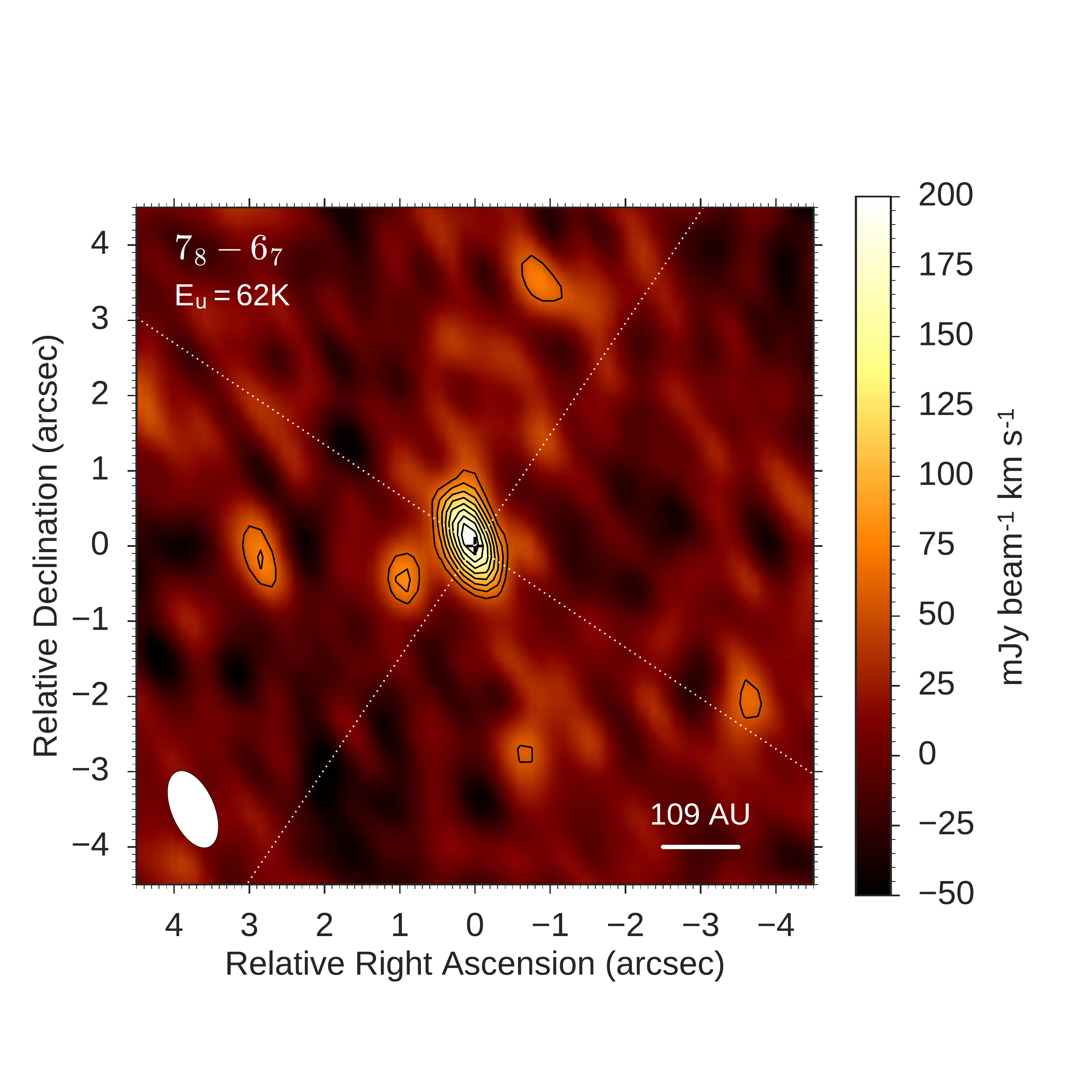}
\caption{Integrated intensity maps of the two SO transitions taken over a 11~km~s$^{-1}$ velocity range.
Left: the $J=7_{7}-6_{6}$ transition with an r.m.s of 22~mJy~beam$^{-1}$~km~s$^{-1}$ and a peak emission of 151~mJy~beam$^{-1}$~km~s$^{-1}$ resulting in a S/N of 6.9.
Right: the $J=7_{8}-6_{7}$ transition with an r.m.s of 19~mJy~beam$^{-1}$~km~s$^{-1}$ and a peak emission of 206~mJy~beam$^{-1}$~km~s$^{-1}$ resulting in a S/N of 10.8.
The black contours are at intervals of $\sigma$ going from 3$\sigma$ to peak.}
\end{figure*}

\begin{figure*}
\resizebox{\hsize}{!}{\includegraphics{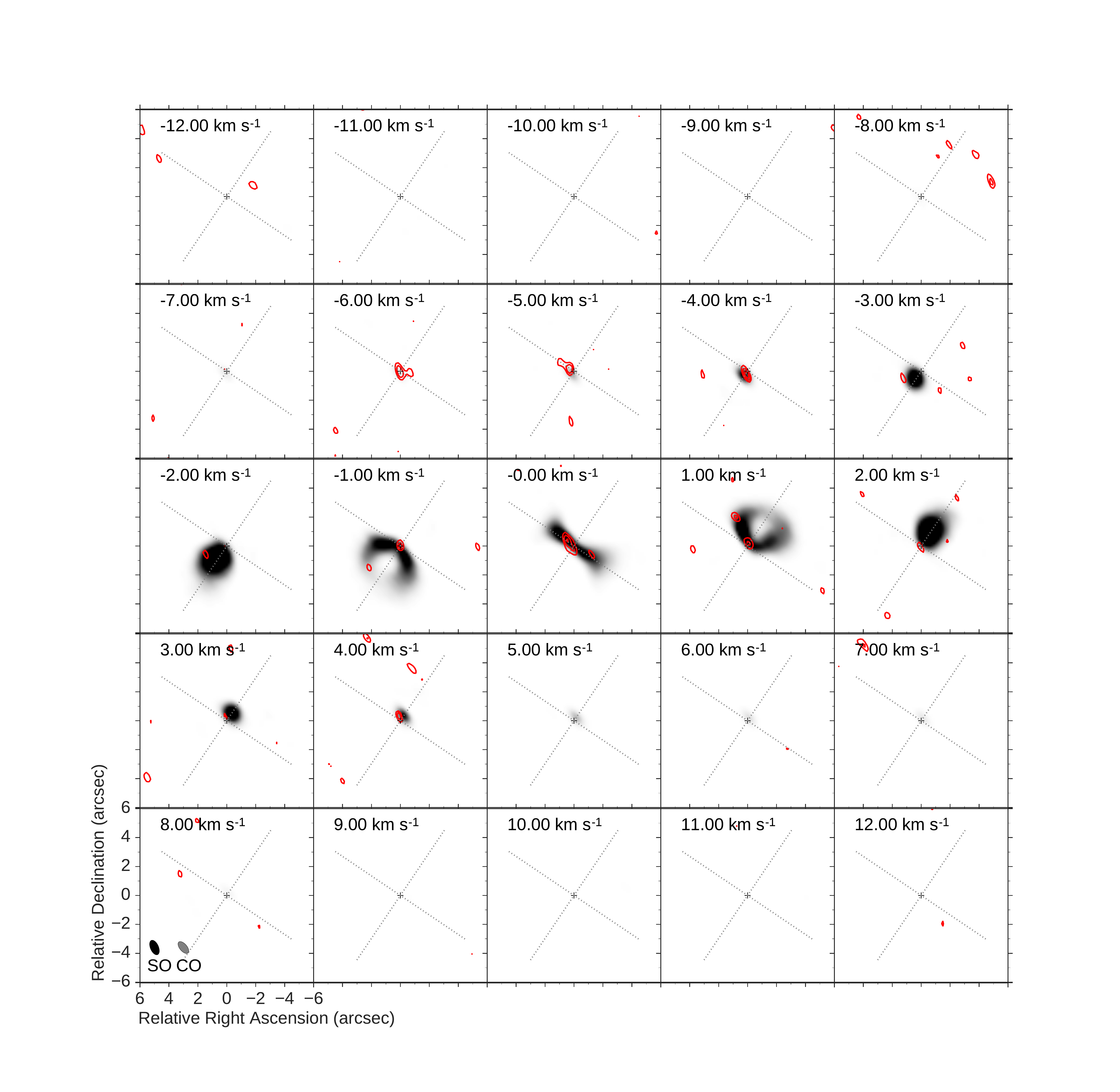}}
\caption{Channel maps of the stacked SO emission (red contours) and the CO $J=3-2$ emission with a 3$\sigma$ clip (grey colour map).
The stacked SO emission has with an r.m.s. noise of 4.2~mJy~beam$^{-1}$ channel$^{-1}$ and peak emission of 24.7~mJy~beam$^{-1}$
resulting in a peak S/N of 5.9.
The contours are from 3$\sigma$ to peak in intervals of $\sigma$. 
The velocities stated are with respect to the source velocity of the emission.}
\end{figure*}

Figure 3 shows the line profile extracted from within the 3$\sigma$ contour of the stacked integrated intensity
and covers a velocity range of $-$~$50$~km~s$^{-1}$ to $+$~$50$~km~s$^{-1}$ about the source velocity. 
This large velocity range was chosen to highlight the significance of the emission with respect to the underlying noise.
The r.m.s. of the line profile was determined from the line-free channels either side of the channels with significant emission.
The line profile is double peaked which could indicate that the emission is originating from an inclined disk in Keplerian rotation. 
However, the trough of the profile is $2$~km~s$^{-1}$ blueshifted from the CO-determined source velocity. 
Because the emission is clearly peaking on source, numerous checks were done to see if the blue-shifted emission shift is real.
The CASA velocity reference frame is LSRK as is needed, the line frequencies are correct, and there are no other emission lines
within the considered velocity range that could attribute to the blue-shifted emission. In addition to this, 
we checked the removal of 5\% to 10\% of the longer and shorter baselines and the double-peaked line profile persists. 
We also checked that the frequency axis of the spectral cube was not incorrectly indexed. 
The line profiles of the individual 
transitions (Appendix A) both have the same line profile shape. Therefore, we did not induce any additional 
signal through stacking.

\begin{figure}
  \centering
  \resizebox{\hsize}{!}{\includegraphics{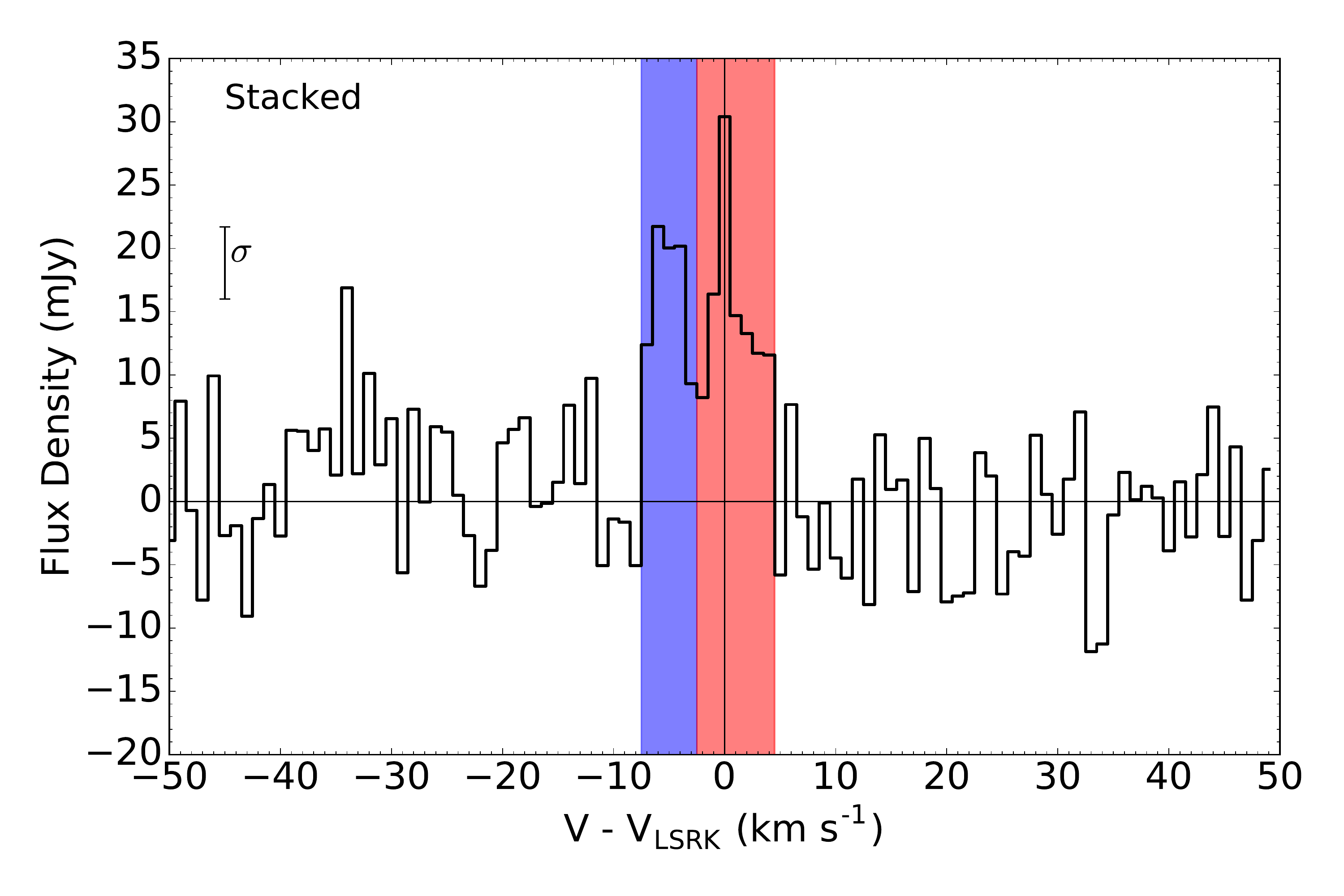}}
  \caption{Line profile extracted from within the 3$\sigma$ extent of the SO stacked integrated intensity with an r.m.s. noise of 5.7~mJy
 and a peak flux of 30.4~mJy resulting in a S/N of 5.3. 
 Highlighted in red and blue are the velocity ranges of emission used in the moment maps of the individual lines in Figure 4.}
  \label{}
\end{figure}

 To investigate the spatial distribution of both components of emission 
 (on source emission and blue shifted emission, respectively), a moment zero map was created for both.
Figure 4 shows the integrated intensity maps from 
-7.5~km~s$^{-1}$ to -2.5~km~s$^{-1}$ and from -2.5~km~s$^{-1}$ to 4.5~km~s$^{-1}$. These velocity ranges are highlighted in the 
stacked line profile (Figure 3) and the individual lines profiles (Appendix A).
The peak S/N and r.m.s. for each of these integrated intensity maps are listed in Table 2.
The peak emission in the two maps is spatially offset but the exact separation of these two components is unclear because the
 emission is of the order of the same size as the beam.
The ratio of the peak emission in the moment zero maps of the $J=7_{7}-6_{6}$ transition to the $J=7_8-6_7$ transition 
 is higher in the blue-shifted component of emission compared with the emission at source velocity. 
 This likely indicates that the blue-shifted component is tracing warmer gas. The stacked moment maps over the two velocity 
 ranges are shown in Appendix C and the spatial offset between the two components is more significant.

\begin{figure*}
\includegraphics[width=0.5\hsize]{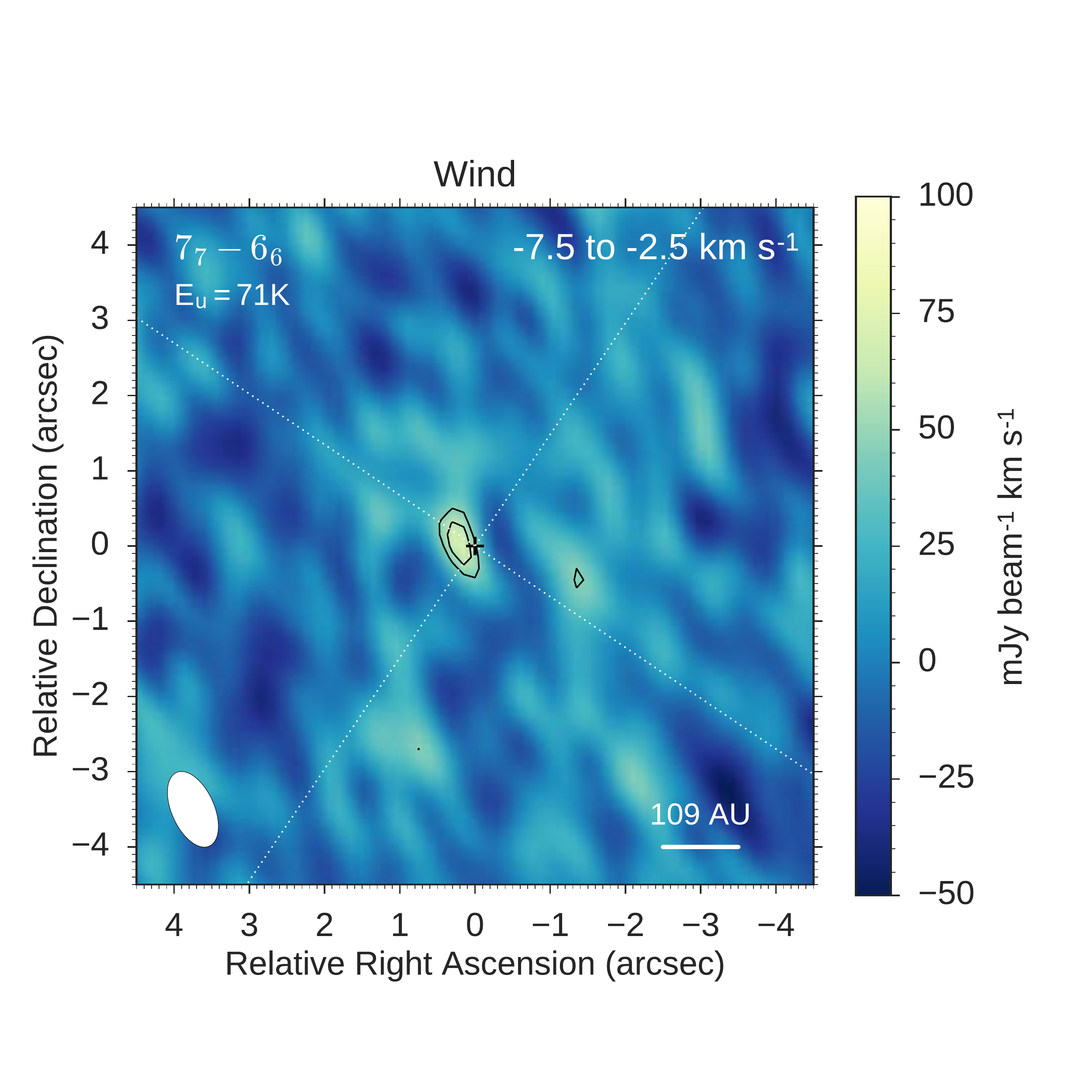}
\includegraphics[width=0.5\hsize]{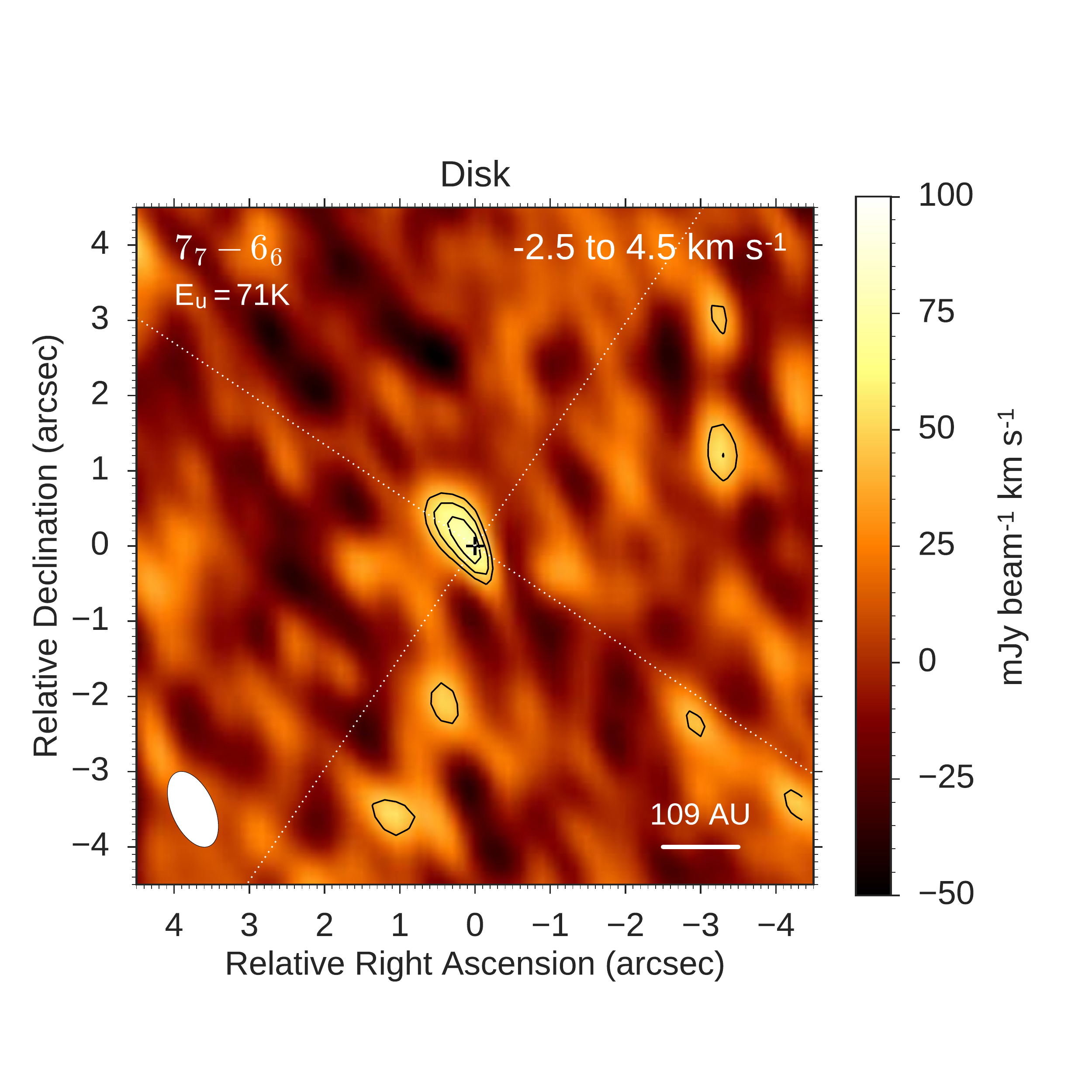}
\includegraphics[width=0.5\hsize]{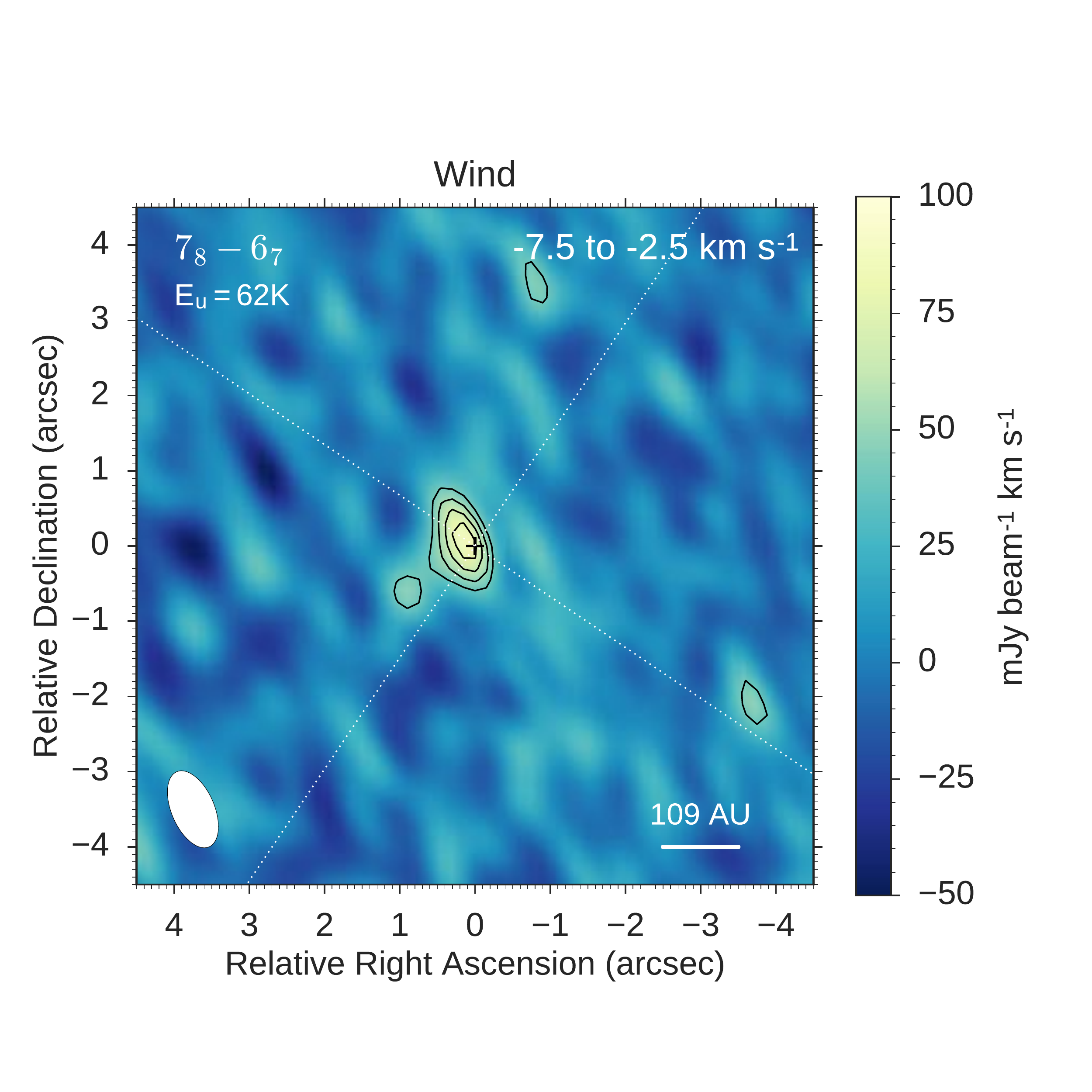}
\includegraphics[width=0.5\hsize]{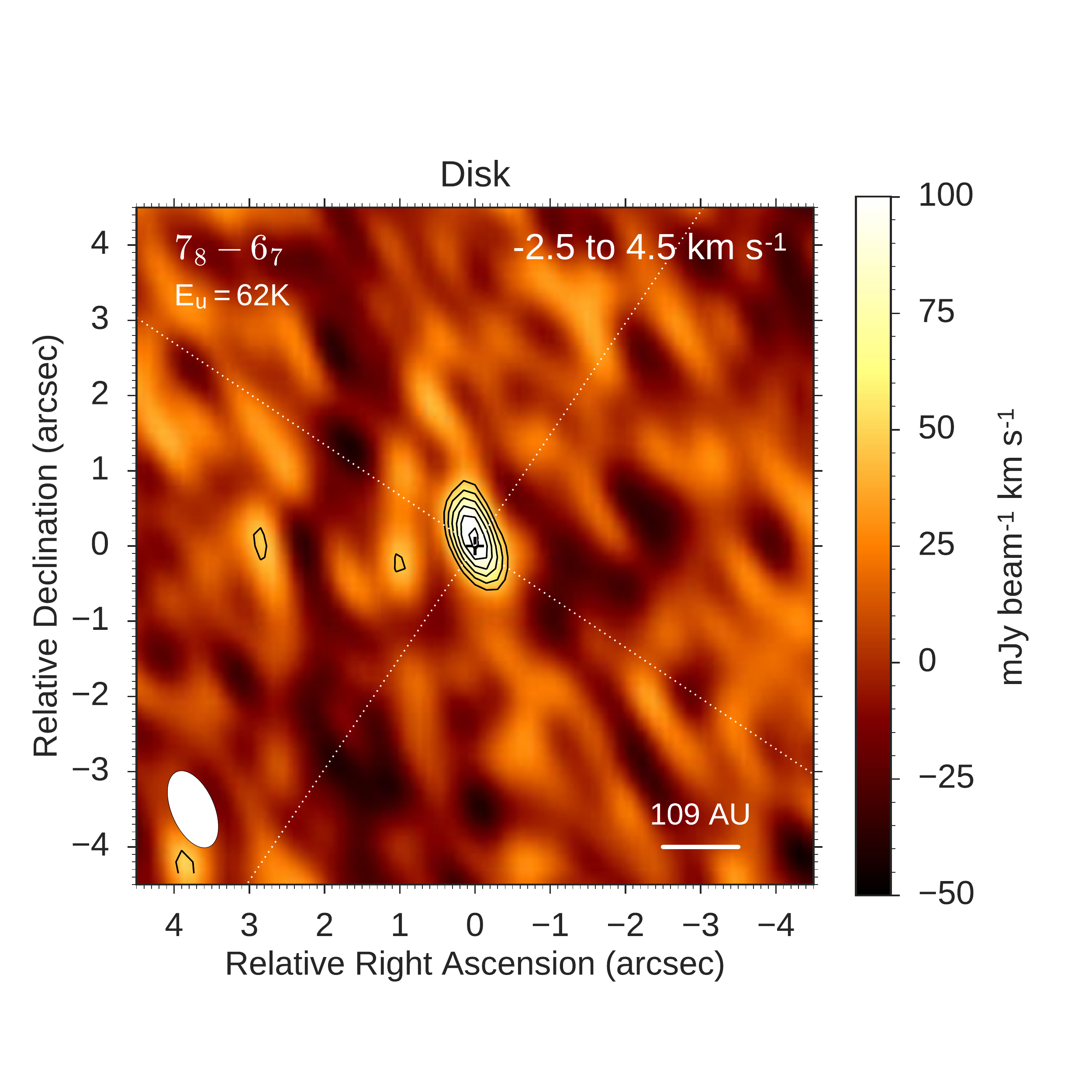}
\caption{
Integrated intensity maps of the two SO transitions over two velocity ranges.
Top: the $J=7_{7}-6_{6}$ transition from -7.5~km~s$^{-1}$to -2.5~km~s$^{-1}$ (left) and -2.5~km~s$^{-1}$to 4.5~km~s$^{-1}$ (right). 
Bottom: the $J=7_{8}-6_{7}$ transition from -7.5~km~s$^{-1}$to -2.5~km~s$^{-1}$ (left) and -2.5~km~s$^{-1}$to 4.5~km~s$^{-1}$ (right). 
The peak S/N and r.m.s. for each of these maps are listed in Table 2.
The black contours are at intervals of $\sigma$ going from 3$\sigma$ to peak.}
\end{figure*}

\begin{table}[]
\centering
\begin{threeparttable}
\caption{Moment map S/N and r.m.s.}
\begin{tabular}{cccc}
\hline\hline
\multicolumn{2}{c}{Moment map}      & \multicolumn{1}{c}{S/N}            & \multicolumn{1}{c}{r.m.s.}                       \\
\multicolumn{2}{c}{}                & \multicolumn{1}{c}{}               & \multicolumn{1}{c}{(mJy~beam$^{-1}$~km~s$^{-1}$)} \\    
\hline
\multicolumn{1}{l}{$J=7_{7}-6_{6}$} & \multicolumn{1}{l}{disk component} & \multicolumn{1}{c}{4.8} & \multicolumn{1}{c}{17} \\
\multicolumn{1}{l}{$J=7_{7}-6_{6}$} & \multicolumn{1}{l}{wind component} & \multicolumn{1}{c}{5.0} & \multicolumn{1}{c}{15} \\
\multicolumn{1}{l}{$J=7_{8}-6_{7}$} & \multicolumn{1}{l}{disk component} & \multicolumn{1}{c}{8.4} & \multicolumn{1}{c}{15} \\
\multicolumn{1}{l}{$J=7_{8}-6_{7}$} & \multicolumn{1}{l}{wind component} & \multicolumn{1}{c}{6.8} & \multicolumn{1}{c}{13} \\
\hline
\end{tabular}
\end{threeparttable}
\end{table}

From this analysis we propose that we are observing two components of emission and not 
just Keplerian disk emission that is blue-shifted relative to the source velocity. 
The emission in the line profile that peaks on source and aligns kinematically with the CO in the
channel maps (Figure 2) can be attributed to disk emission. 
The morphology of the on-source singly-peaked disk component might be explained by asymmetric SO emission which 
is arising from the north-east side of the disk only. This is investigated further in Section 5.
The blue-shifted emission peaks off source spectrally and spatially and it is attributed to a disk wind
where material is being driven from the surface of the disk resulting in a blue-shift along the line of sight.

\subsection{Confirmation of the SO detection via matched filter analysis}

We use a matched filter code\footnote{vis\_sample is publicly available at: \hfill \break \url{https://github.com/AstroChem/vis\_sample}}
developed for interferometric data sets to
confirm the detection of the two SO lines and 
to search for emission from the lines undetected in the image plane \citep{loomis2017}.
The detection of weak spectral lines direct from the \textit{uv} data is optimal as
this saves the computational time needed to generate images and eliminates imaging biases. 
Since the position-velocity pattern of a disk in Keplerian rotation is well characterised, matched 
filtering can be used to detect weak spectral lines in disks \citep{loomis2017}.
This technique has been shown to increase S/N of the methanol detection in TW Hya by 53$\%$ \citep{Walsh2016}.
 A filter can either be a strong line detected in the same disk or a model of the anticipated 
emission.
The image cube used as a filter is sampled in the \textit{uv} plane and these visibilities
are cross-correlated with the low S/N visibilities. This is done by sliding the filter though the data channel-by-channel 
along the velocity axis. If there is a detectable signal, i.e. emission with a similar position and
velocity distribution as the filter, the filter response will peak at the source velocity of the emission.

We use the CLEAN image of the CO line ($J=3-2$) and the best fit LIME model
(see Section 5 for details) as filters. 
The CO image has been scaled down to one quarter of its size and 
this is motivated by the compact nature of the SO emission 
(as seen in the channel maps and the integrated intensity: Figures 1 and 2). Application of the matched filter results in a 
measure of the response of the filter to the data at
a given frequency. The response is scaled to $\sigma$ with the r.m.s. noise normalised to 1.

Figure 4 shows the response of the three detected SO lines to the compact CO and best fit LIME wedge model filters.
For the CO filter three of the four lines were detected, the $J=7_8-6_7$, $J=7_{7}-6_{6}$ and $J=8_{8}-7_{7}$ transitions.
They have a peak response of 6.8, 4.2 and 4.0 respectively. 
There is no detection of the lower energy J=$3_2-1_2$ transition. 
We tested different CO filters with different compression factors of 1/2, 1/3 and 1/4.
We found that the $J=7_8-6_7$ transition is picked up in all the filters but
the higher excitation lines have an improved response with the more
compact filters. The best fit LIME model also detects the same three SO lines. 
The $J=8_{8}-7_{7}$ filter response is approximately the same as with the 
compact CO filter but the other two lines responses are quite different.

The peak of the responses are not all at the expected source velocity
supporting the theory that we have multiple velocity components of emission including a possible disk wind.
The matched filter has confirmed the detection of the two lines detected in the image plane
and they are observed at a substantially higher S/N than in the channel maps. It has also facilitated the detection of a 
line that we do not detect in the imaging. Further, the matched filter line response scales
with intensity, so we also now have rudimentary excitation information
to motivate our models.

\begin{figure*}
\begin{subfigure}{.5\hsize}
\centering
\includegraphics[width=0.95\hsize]{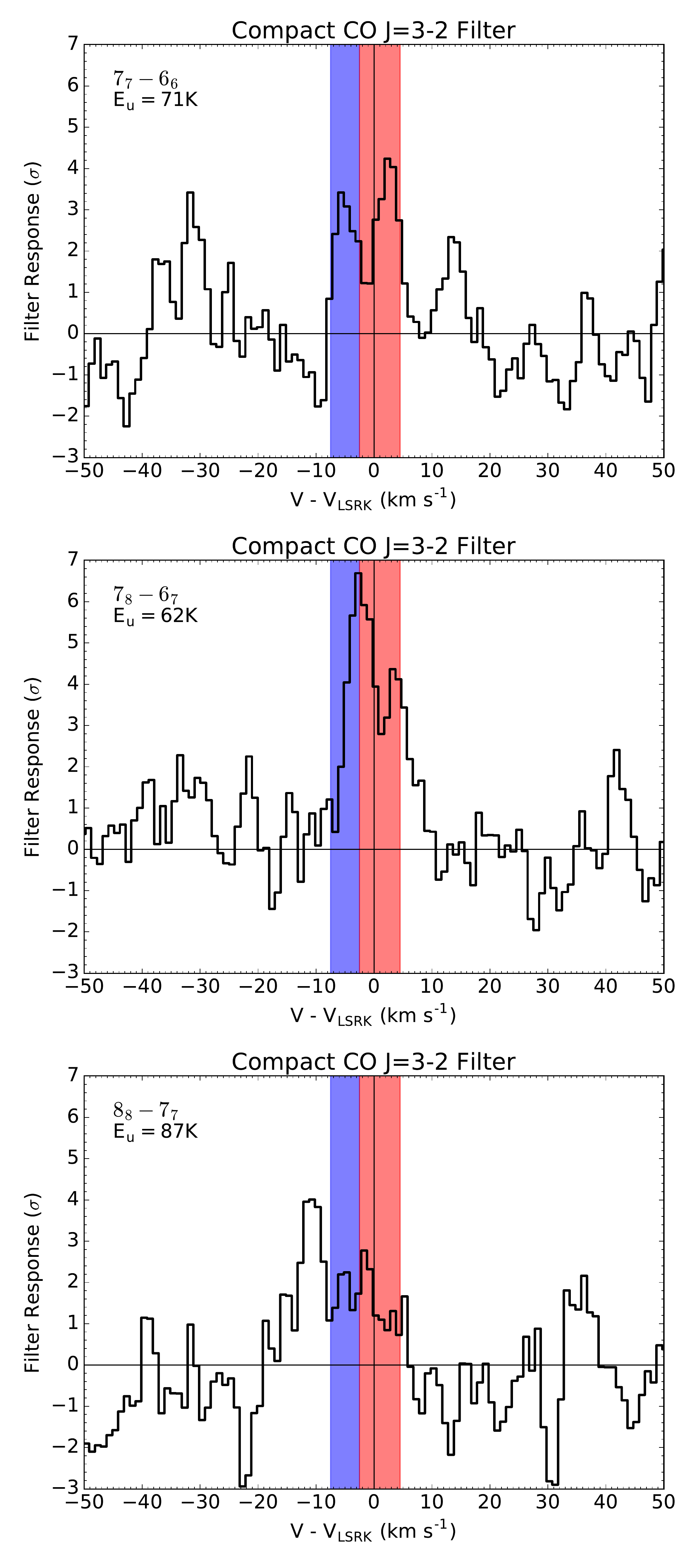}
\label{}
\end{subfigure}
\begin{subfigure}{.5\hsize}
\includegraphics[width=0.95\hsize]{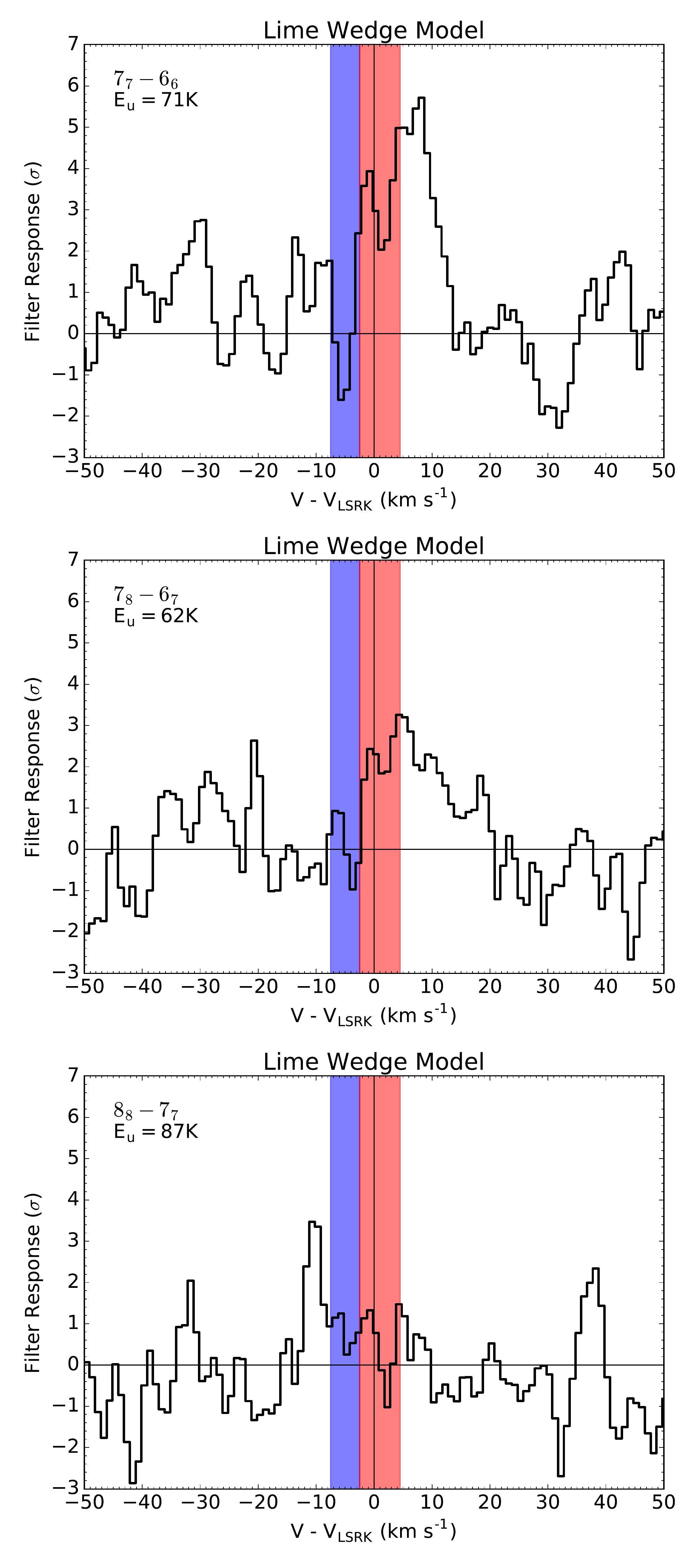}
\label{}
\end{subfigure}
\caption{Matched filter responses for the three detected SO transitions. Left:
the results using a spatially compact (1/4) version of the CO channel maps as a filter.
Right: the results from using the best fit LIME wedge model as a filter (see Section 5 for details).
Highlighted in red and blue are the velocity ranges of emission we attribute to a disk and a wind component respectively.}
\end{figure*}

\section{Modelling the SO disk emission using LIME}

The abundance of SO was estimated by matching the observed emission with simulated emission generated using 
a HD 100456 physical disk structure (Figure 6)
and LIME version 1.5 \citep[LIne Modelling Engine;][]{2010A&A...523A..25B}. 
Ray-tracing calculations were done assuming LTE and the appropriate distance, inclination and position angle for the source.
The molecular data files for sulphur monoxide were taken from the Leiden Atomic and
Molecular Database (LAMDA; http://home.strw.leidenuniv.nl/$~$moldata/SO.html).
To check that LTE calculations were a good approximation we calculated the critical density of the 
transitions.
When the number density of gas is greater than the critical density collisional processes dominate.
In this regime the level populations are determined by the Boltzmann distribution and LTE is an 
accurate assumption. 
The critical density is determined by:
$$ n_{cr} = \frac{A_{ul}}{\sum\limits_{l'<u}{\gamma_{ul'}}} $$
where $A_{ul}$ and $\gamma_{ul}$ are the Einstein A and collisional rate coefficients of the transition.
The critical density for the SO $J=7_{7}-6_{6}$ transition was determined to be $6~\times~10^{6}$~cm$^{-3}$ at 100~K using the LAMDA molecular
data with the collisional rates from \cite{2006A&A...450..399L}. The other transitions are of a similar order of magnitude. 

As the matched filter is a linear process the relative responses for a pair of lines can be used
as a proxy for their relative intensities after correcting for the difference in noise levels in each line. 
These relative intensities were converted to line ratios and this information was used to confine the location of the SO in the disk with respect to the temperature and density conditions.
Model line ratios were calculated from line intensities determined using the RADEX radiative transfer code assuming an SO column density of 10$^{14}$~cm$^{-2}$ motivated by full chemical models. 
RADEX is a non-LTE 1D radiative transfer code that can be used 
with the intensity of an observed particular molecular line to estimate the excitation temperature and column density of the gas,
assuming an isothermal, homogeneous medium with no significant velocity gradient \citep{vanderTak:2007be}.
The line ratios of the three lines were calculated over a grid of temperatures and densities and the 
results are shown in Figure 7. These model line ratios were compared to the observed line ratios taken from the matched filter responses.
Within the velocity range we defined as disk emission, these are
1.5, 1.6 and 1.1 for the $J=7_{8}-6_{7}$/$J=7_{7}-6_{6}$, $J=7_{8}-6_{7}$/$J=8_{7}-7_{6}$ and $J=7_{7}-6_{6}$/$J=8_{7}-7_{6}$ 
line ratios respectively. A selection of the RADEX results are shown in a table in Appendix D along with the observed ratios and their associated errors.
The regime that best fits our observations is a  H$_2$ density between 10$^8$ to 10$^{10}$~cm$^{-3}$ and a gas
temperature between 50 and 100~K. These conditions result in the SO being distributed in a ring from 20 to 100~au in a layer above the mid-plane (see Figure 6).
This is in agreement with previous modelling of sulphur volatiles in disks \citep[e.g.][]{Dutrey2011} and is in agreement with the
compact nature of the SO emission we observe. 
Modelling the SO in a region of lower density and higher temperature resulted in significantly more 
extended emission than in our observations.
We opt to only model the near surface of the disk as we assume that in this region of the disk (<100~au) the optically thick dust emission 
will block the emission from the molecular gas in the far surface of the disk. 

\begin{figure*}
\begin{subfigure}{.5\hsize}
\includegraphics[width=\hsize]{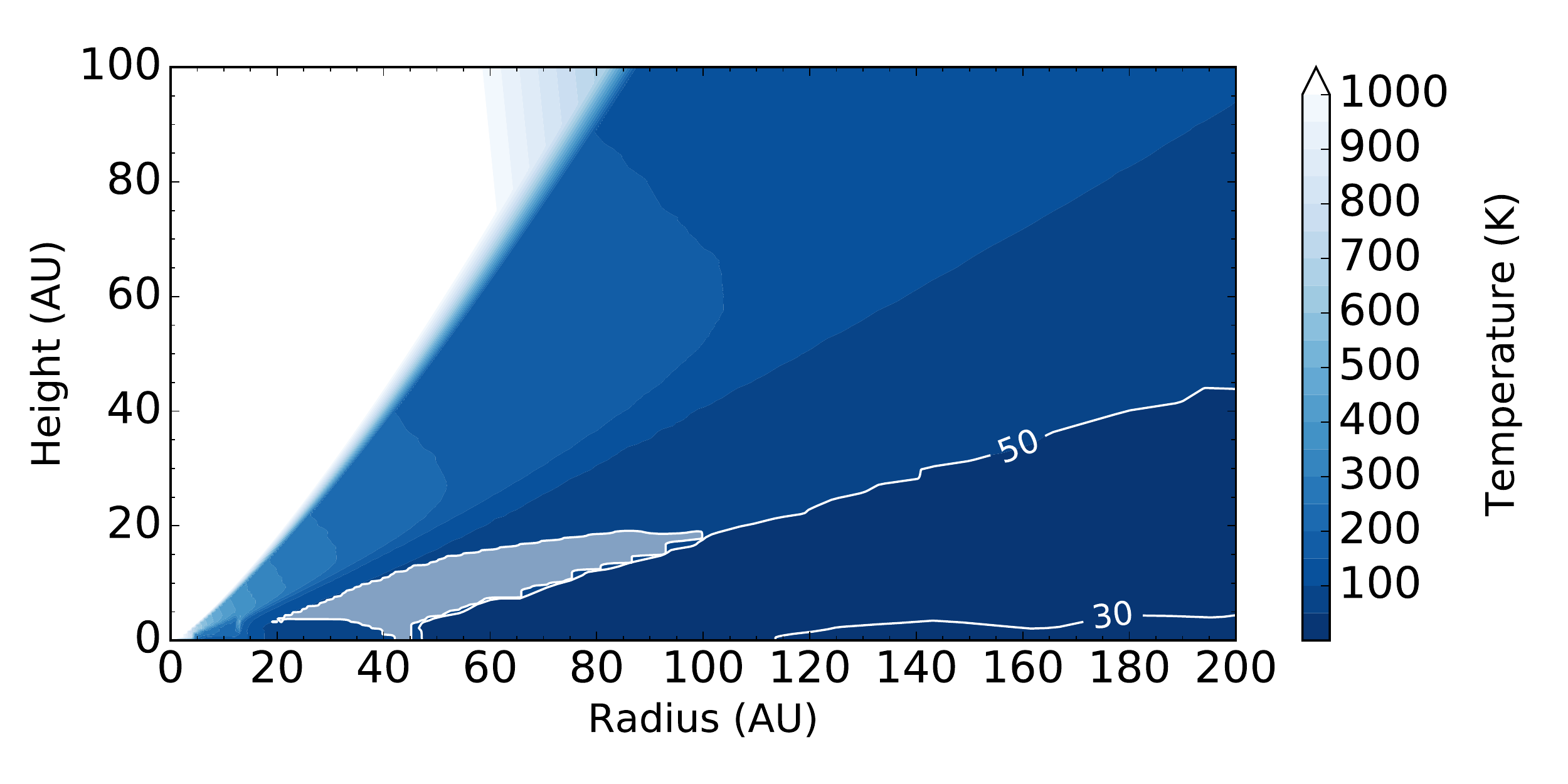}
\label{}
\end{subfigure}
\begin{subfigure}{.5\hsize}
\includegraphics[width=\hsize]{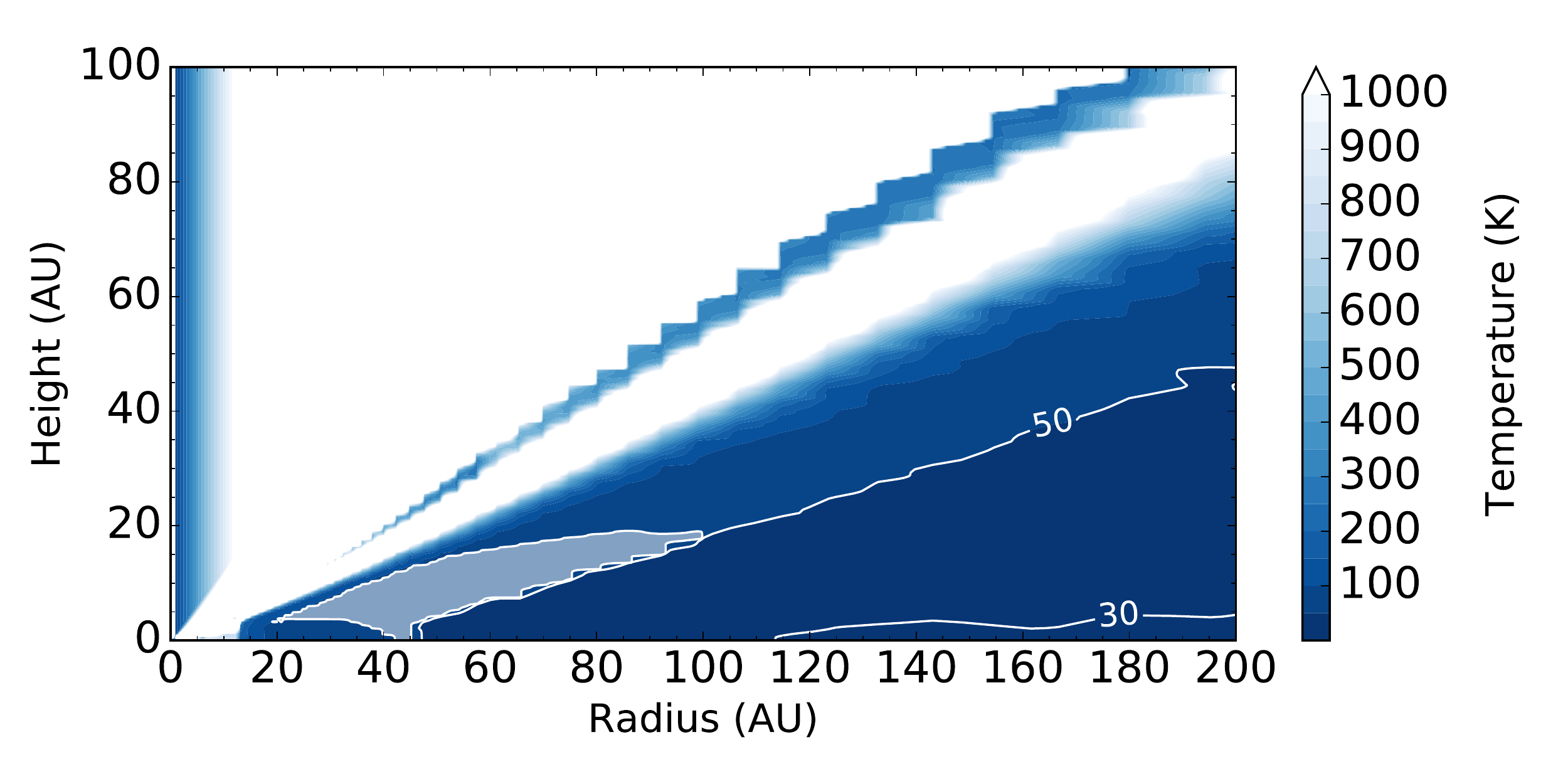}
\label{}
\end{subfigure}
\begin{subfigure}{.5\hsize}
\includegraphics[width=\hsize]{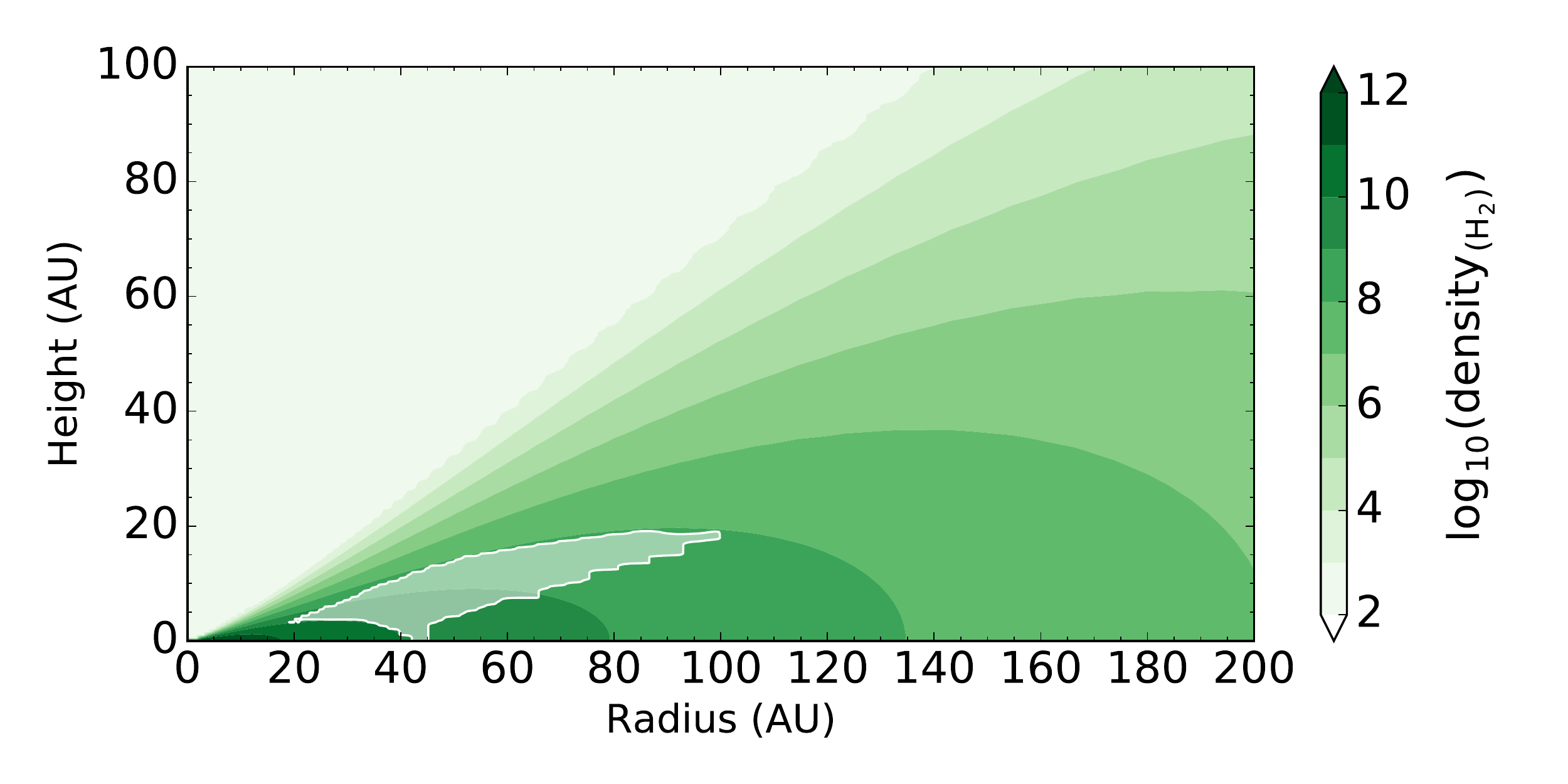}
\label{}
\end{subfigure}
\begin{subfigure}{.5\hsize}
\includegraphics[width=\hsize]{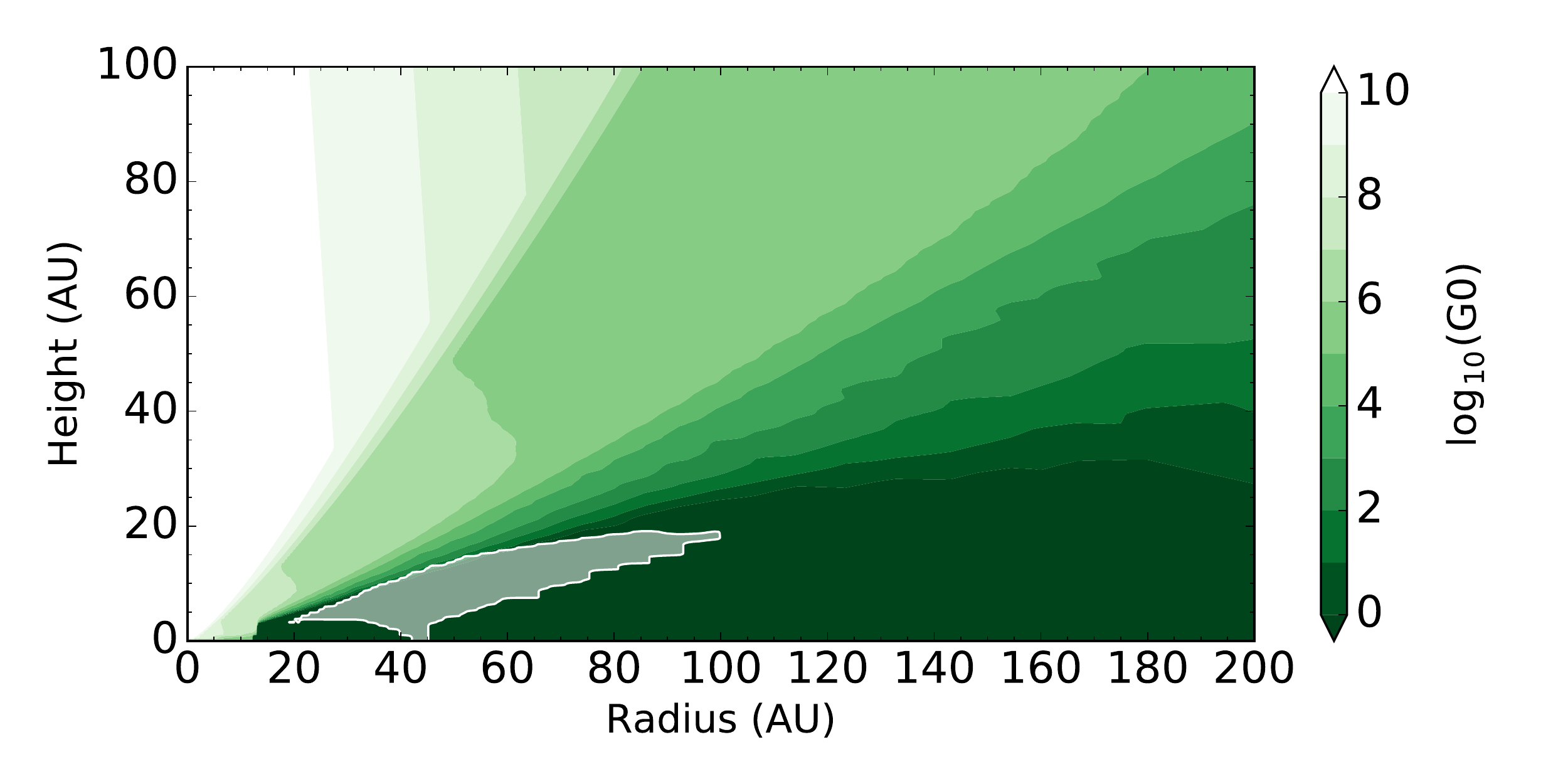}
\label{}
\end{subfigure}
\caption{The HD 100546 disk physical structure from \citet{Kama2016}.  Top left and moving clockwise:
 the dust temperature (K), gas temperature (K), UV flux (in units of the interstellar radiation field)
  and number density (cm$^{-3}$). The two white contours in each of the temperature plots 
  correspond to temperatures 30~K and 50~K. The shaded region highlights the location of the SO 
  motivated by RADEX calculations and used in the LIME modelling.}
  
\end{figure*}

\begin{figure*}
\centering
\includegraphics[width=0.95\hsize]{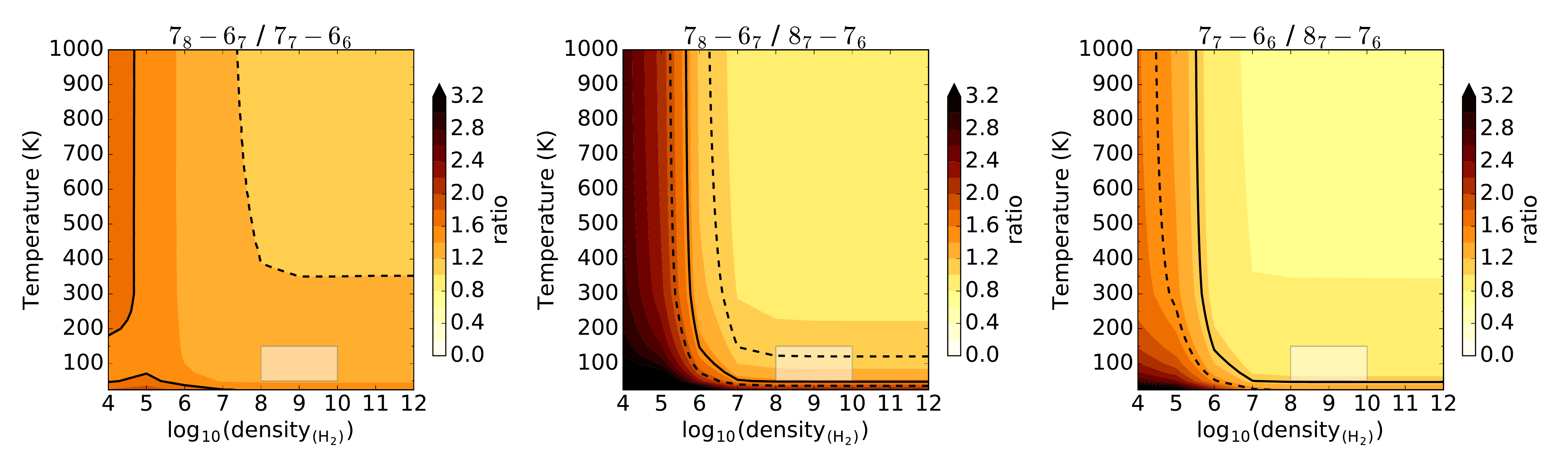}
\caption{The RADEX modelling results of the selected SO line ratios. From left to right are the three ratios:
$J=7_{8}-6_{7}$/$J=7_{7}-6_{6}$, $J=7_{8}-6_{7}$/$J=8_{7}-7_{6}$ and $J=7_{7}-6_{6}$/$J=8_{7}-7_{6}$. 
The solid black line is the observed line ratio and the dotted black lines are the error bars. 
The shaded region highlights the temperature and density conditions chosen for the location of the SO in the LIME modelling. }
\end{figure*}

A set of models were run varying the fractional abundance of SO with respect to H$_2$. This was done in order to 
match the observed peak in the integrated intensity of the disk component for each of the two transitions:
80~mJy~beam$^{-1}$~km~s$^{-1}$ for the $J=7_{7}-6_{6}$ transition and 124~mJy~beam$^{-1}$~km~s$^{-1}$ for the $J=7_8-6_7$. 
A model for a full disk was calculated with a fractional abundance of 3.5$\times$~$10^{-7}$ 
with respect to H$_2$ resulting in a peak intensity for each of the lines
of 92~mJy~beam$^{-1}$~km~s$^{-1}$ and 109~mJy~beam$^{-1}$~km~s$^{-1}$ respectively.
These values match the peak emission of the observations within the 1$\sigma$ range.
The residual maps of the observed integrated intensity minus the model integrated intensity for the two lines are shown in Figure 8. 
The residuals show that the observed emission is asymmetric peaking in the north east region of the
disk and a full disk is not an accurate representation of the data. 
A second model was run restricting the SO to a specific angular region of the disk.
A 45\degree~wedge of emission was calculated with the optimal position picked by eye from the residual
maps to be from 0\degree~ to 45\degree~ from the disk's major axis.
A fractional abundance of 5.0$\times$~$10^{-6}$ with respect to H$_2$
resulted in a peak intensity for each of the lines of 93~mJy~beam$^{-1}$~km~s$^{-1}$ 
and 96~mJy~beam$^{-1}$~km~s$^{-1}$ respectively.  
These values match the peak emission of the observations within the 2$\sigma$ range.
This fractional abundance is greater than the `depleted' sulphur fractional abundance 
observed dark clouds (SO~$\approx~10^{-8}$; \citealt{Ruffle1999}). This suggests that 
there are energetic processes occurring in HD100546 releasing a source of refractory sulphur 
into the gas phase. 
The fractional abundance of SO derived from the LIME modelling is model dependent as it 
depends on the gas density of the region of the disk where the SO is located. 
This model well reproduces the integrated intensity however, the kinematics trace red-shifted disk emission.
The peak in both of the line profiles for the wedge models is $1.7\times$ the observed line profile peaks and the 
model emission is over a narrower velocity range.  
The model line profiles for both the disk and the wedge models are compared with the observed line profiles in Appendix E.
Further refinement of the disk emission component requires better data as the emission is the same size scale as the beam.

\begin{figure*}
\includegraphics[width=0.5\hsize]{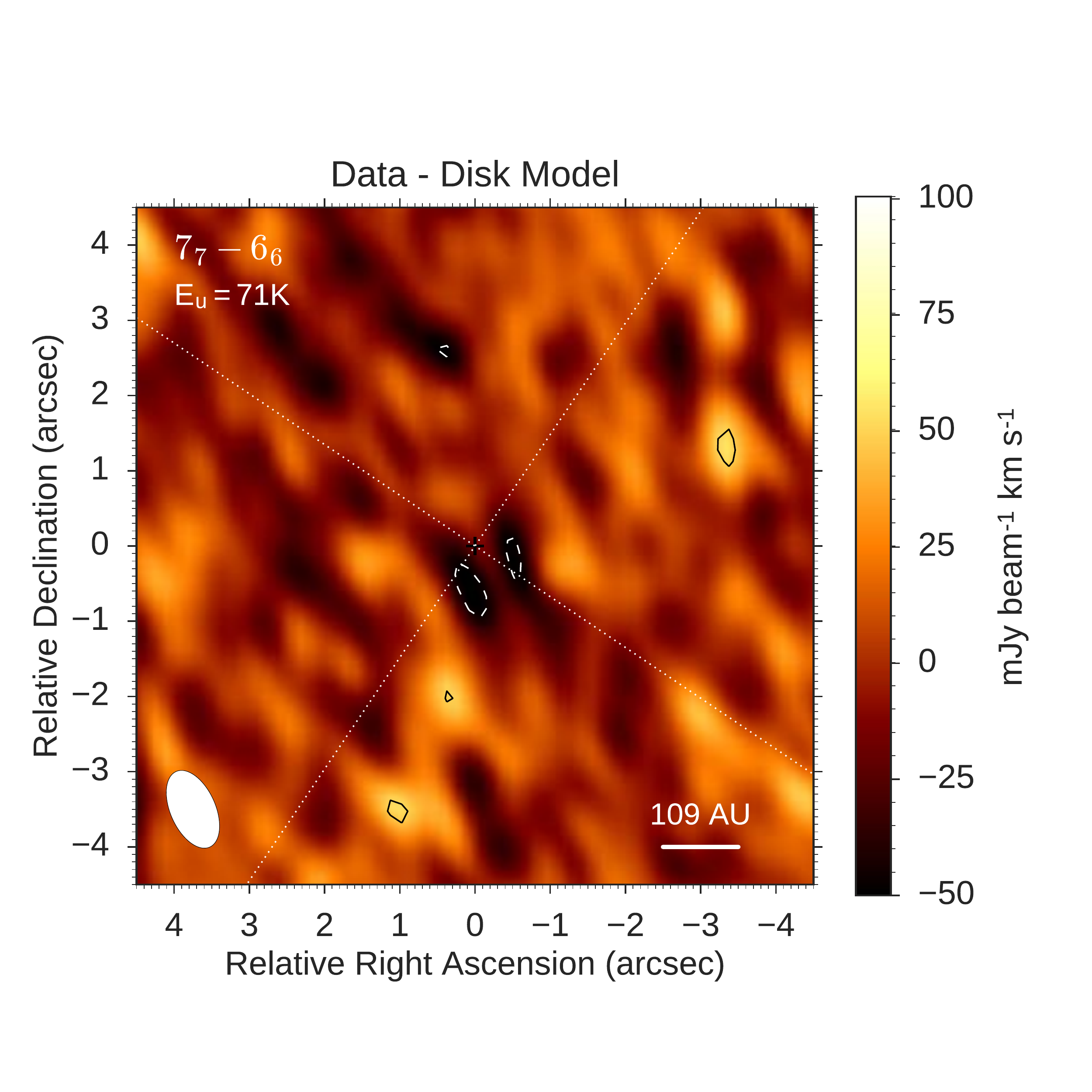}
\includegraphics[width=0.5\hsize]{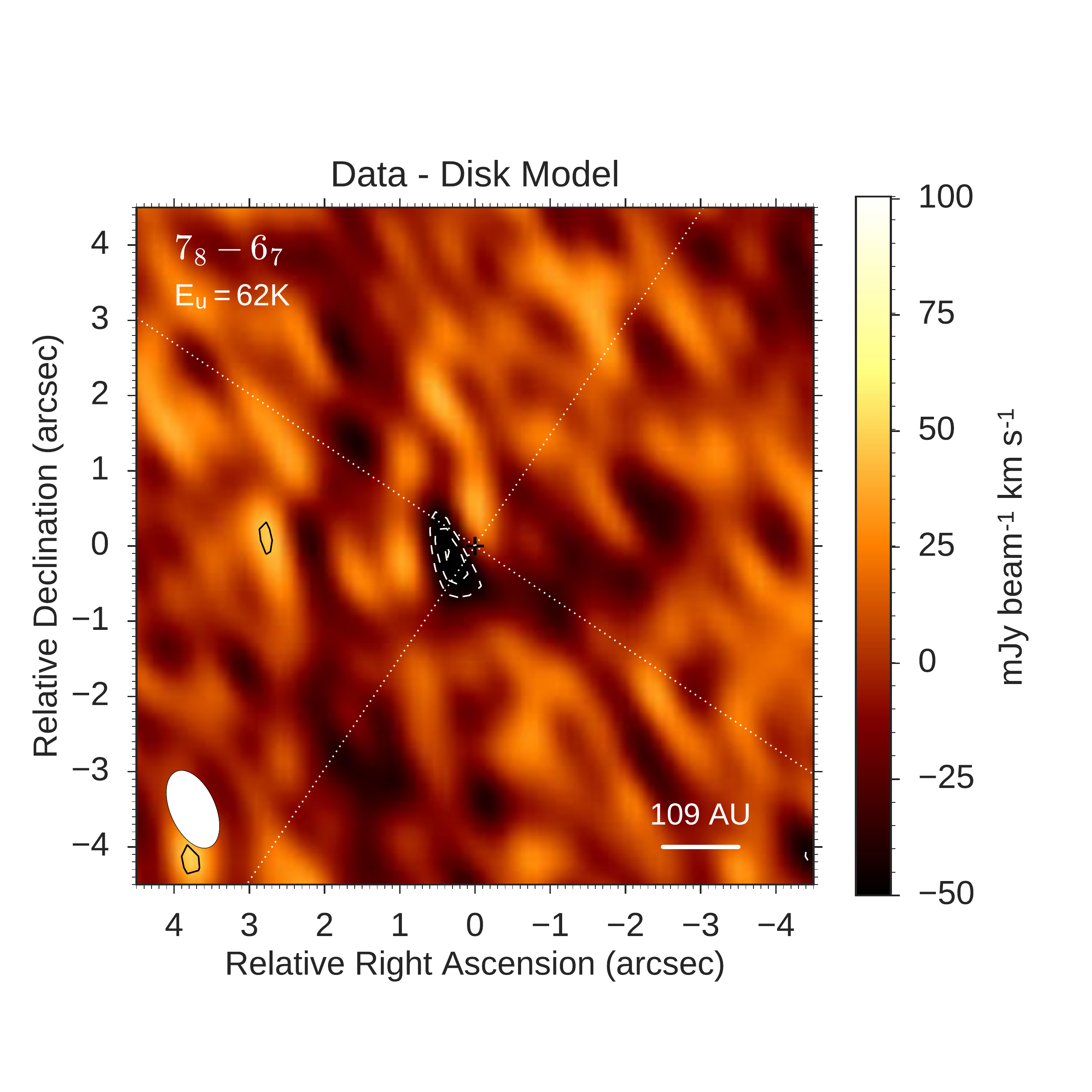}
\includegraphics[width=0.5\hsize]{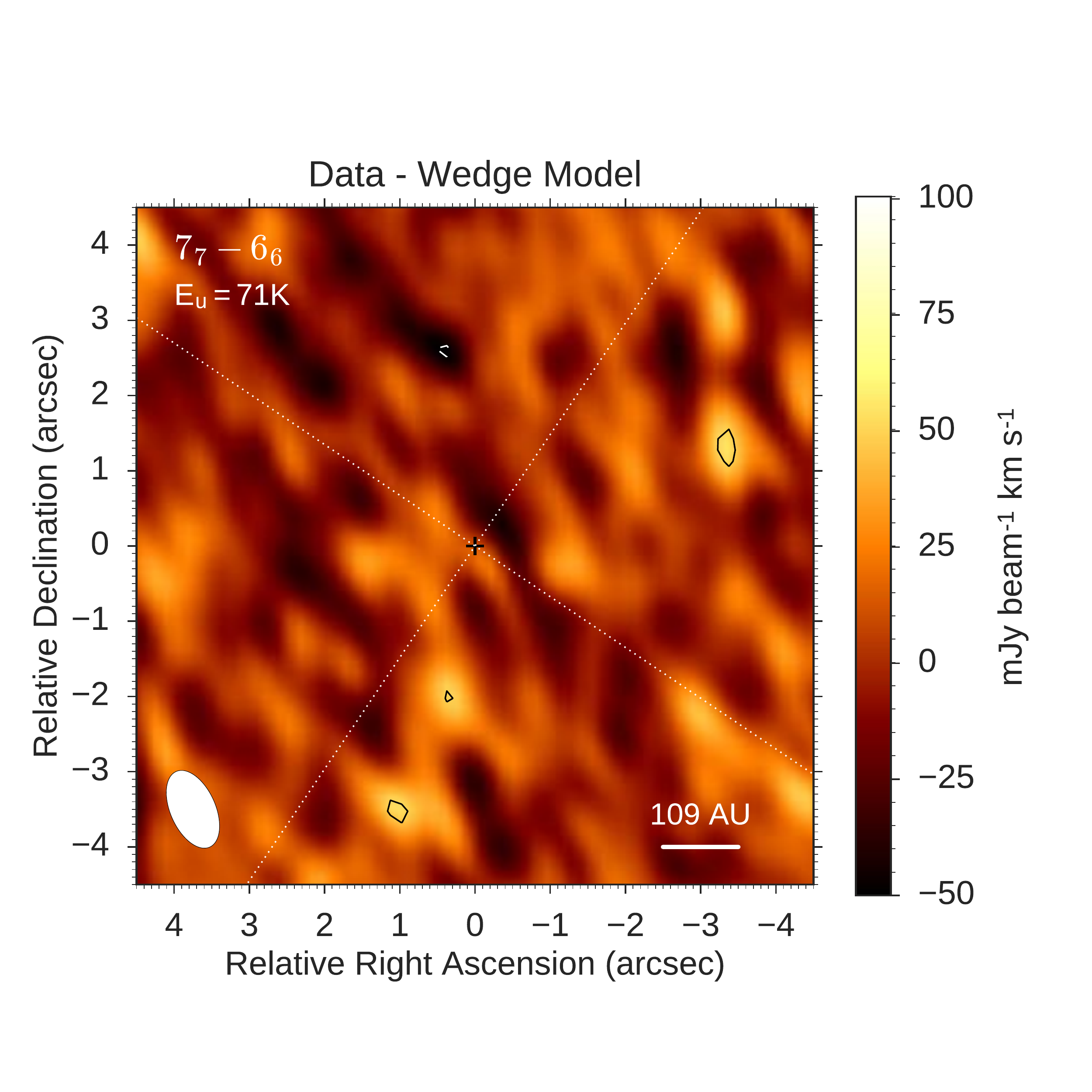}
\includegraphics[width=0.5\hsize]{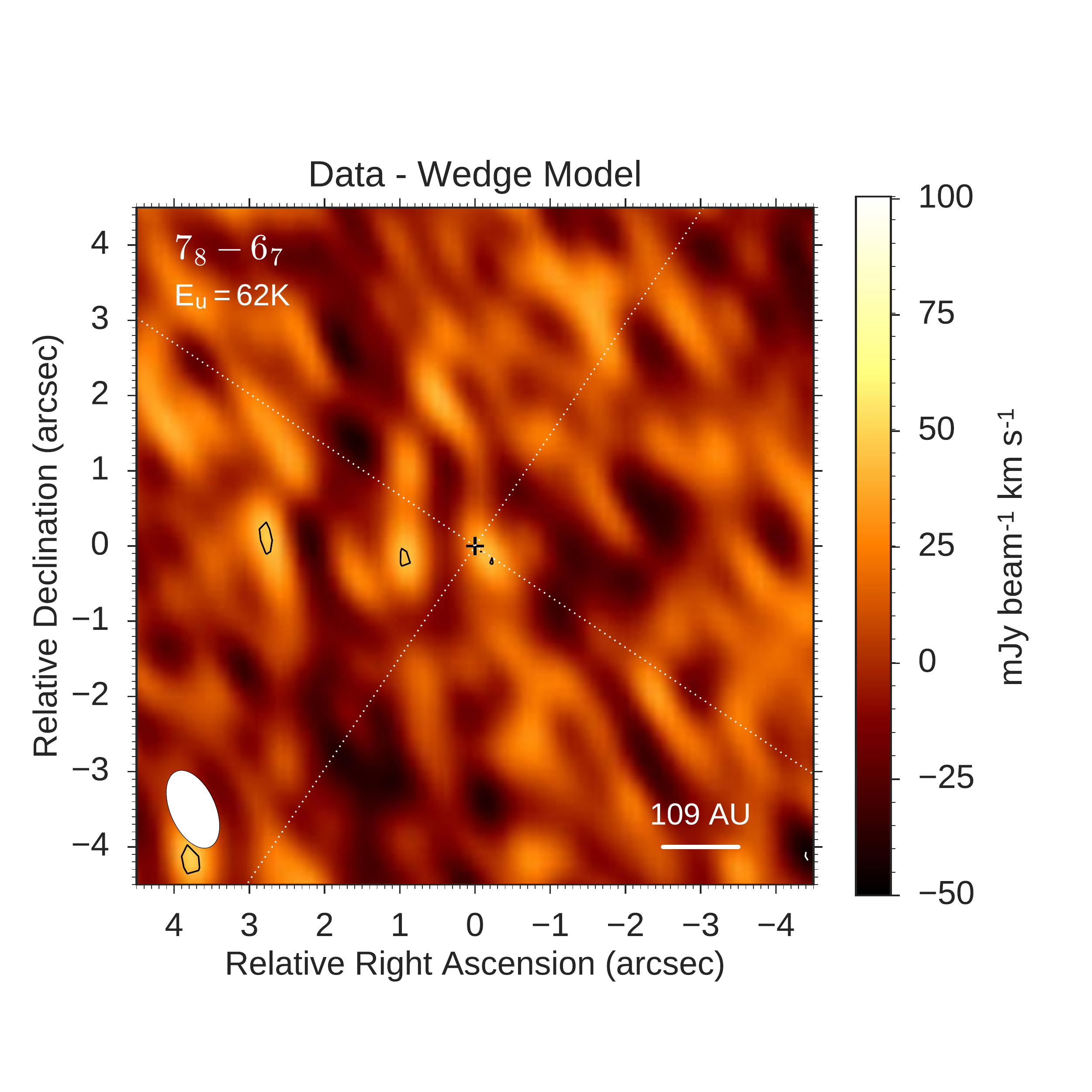}
\caption{Residual maps from the disk emission integrated intensity and the LIME models for each of the transitions. 
Top: disk emission minus disk model. Bottom: disk emission minus wedge model. Overlaid are 
dashed -5,-4 and -3$\sigma$ contours and solid 3, 4 and 5$\sigma$ contours.}
\end{figure*}

\section{Discussion}
\subsection{Location and abundance of the detected SO emission}

We detect SO in the protoplanetary disk around HD 100546 for the first time.
In the image plane we have a clear detection of two lines in the 
integrated intensity maps and we improve the S/N in the channel maps and line profile by stacking.
From the morphology of the line profile and the asymmetric distribution 
of the emission it is likely we are observing two components of emission: 
a wedge of disk emission and a blue-shifted component (-5~km~s$^{-1}$). 
We use a matched filter to better determine the relative intensities of the 
SO lines in the data set and confirm the detection of three transitions and a
non detection of the lowest energy transition. The relative intensities of the three detected lines are used to motivate
the location of SO in LIME modelling.
The residuals from the observed and modelled integrated intensity maps reveal that the 
integrated emission is indeed asymmetric peaking north-east of the source position.
This is coincident with a `hot-spot' observed in CO emission relating to a possible disk warp \citep{walsh2017}.
The CO $J=3-2$ emission from the HD100546 disk is asymmetric along the minor axis with the
emission peaking in the north east region of the disk.
Since this CO emission is optically thick it should be tracing the temperature of the gas
and therefore reflects an non-axisymmetric temperature structure. 
The SO can be modelled as a wedge of emission in this region. 
The excess blue-shifted (-5~km~s$^{-1}$ with respect to the source velocity) component is spatially 
inconsistent with the expected location of blue-shifted Keplerian disk emission.
We attribute this emission to a disk wind. This hypothesis is summarised in a cartoon in Figure 9.

\begin{figure}[h!]
\centering
\includegraphics[trim={3cm 9cm 3cm 9cm},clip, width=\hsize]{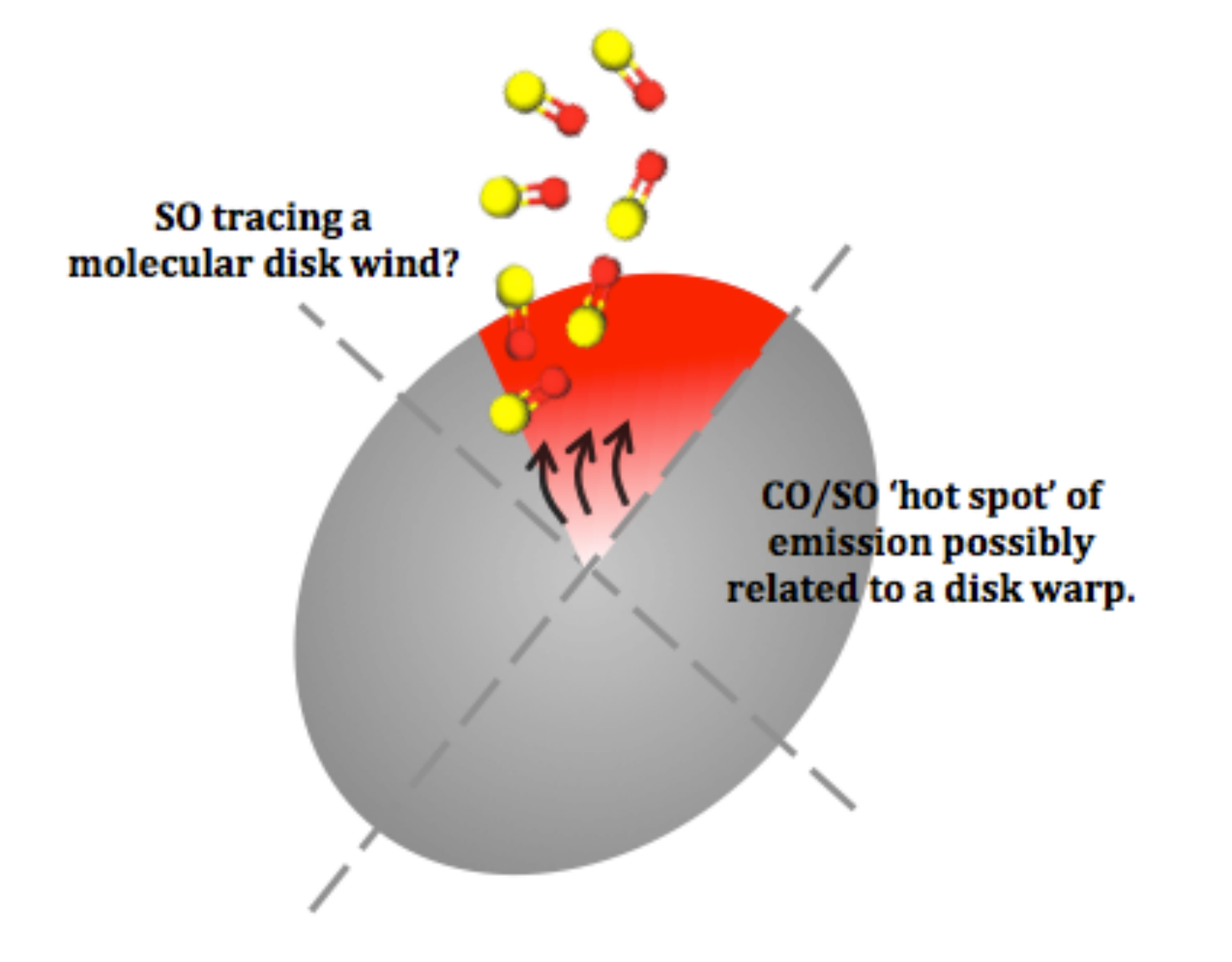}
\caption{Cartoon of HD 100546 wedge disk emission and SO disk wind.}
\end{figure}

We did not detect any of the four SO lines in the complementary HD 97048 Cycle 0 data \citep[see][]{2016Walsh}
using the imaging methods detailed in Sections 3 and 4. The stacked emission in 
the channel maps at a velocity resolution of 1~km~s$^{-1}$ reaches an r.m.s. noise of 6~mJy beam$^{-1}$, and there was no significant response using the matched filter analysis.
This supports our hypothesis that SO is tracing a physical mechanism unique to HD 100546. 
The disks around HD 100546 and HD 97048 have significantly different structures:
the gap(s) in the sub-mm dust are further from the star in HD 97048 and there are no 
protoplanet candidates yet detected in this disk \citep[][]{2012A&A...538A..92Q, 2016Walsh, 2017Plas}.
Further detailed modelling is required to determine the chemical origin of the SO emission in the 
HD 100546 disk and how the physical and chemical conditions differ from those in the HD 97048 disk. 

The only other disk from which SO emission has been imaged is AB Aur \citep{Pacheco2016}. 
In this transition disk the SO is located further from the star in a ring
from approximately 145 to 384~au with a maximum modelled abundance of 2$\times$$10^{-10}$ with respect to H$_2$.
The SO, like in HD 100456, is thought to reside in a layer between the surface and the midplane of the disk.
The relative abundance of SO observed in AB Aur is a few orders of
magnitude less than in HD 100546 and the emission is not necessarily tracing the same 
process. In AB Aur, SO is proposed as a chemical tracer of the early stages
of planet formation as the abundance of SO appears to decrease towards
the disk's dust trap, a local pressure maximum thought to be 
the site of future plant formation \citep[e.g.][]{2013Sci...340.1199V}.

\subsection{Sulphur chemistry in the HD 100546 disk}

Sulphur chemistry, particularly the evolution of S-bearing 
molecules on grain surfaces, is not fully understood as current models
fail to reproduce observed abundances \citep[e.g.][]{Guilloteau2016}.
If we are observing a wedge or partial ring of SO, the inner edge of the emission coincides with the inner edge of the sub-millimetre dust ring at approximately 20~au.
Since HD 100546 is a transition disk the midplane material is exposed to far-UV photons
from the central star. This will cause the desorption of molecules from icy grain mantles.
$H_2O$ ice has been observed in this disk \citep{2016ApJ...821....2H} and $H_2S$ ice is 
a primary component of cometary ices \citep{Bockel2000}. The photodissociation of these molecules
originating from cosmic rays or UV photons, depending on the height of the gas in the disk, would create the 
reactants possible to form SO;
\begin{align}
\ce{H2O} + h\nu &\longrightarrow \ce{OH} + \ce{H} \nonumber \\
&\longrightarrow \ce{O} + \ce{H} + \ce{H} \nonumber \\
\ce{H2S} + h\nu &\longrightarrow \ce{HS} + \ce{H} \nonumber \\
&\longrightarrow \ce{S} + \ce{H} + \ce{H} \nonumber \\
\ce{HS} + \ce{O} &\longrightarrow \ce{SO} + \ce{H} \nonumber \\
\ce{S} + \ce{OH} &\longrightarrow \ce{SO} + \ce{H}.  \nonumber
\end{align}
The activation energy, $E_A$, of the two SO formation reactions is zero (KIDA: KInetic Database for Astrochemistry http://kida.obs.u-bordeaux1.fr/). 
There are a few possible reactions for the 
destruction of SO to form $SO_2$ \citep{1990A&A...231..466M};
\begin{align}
\ce{SO} + \ce{O} &\longrightarrow \ce{SO2} + h\nu  \nonumber   \\
\ce{SO} + \ce{OH} &\longrightarrow \ce{SO2} + \ce{H}. \nonumber 
\end{align}

In AB Aur the abundance of SO decreases with increasing density towards the disk's dust trap. This is attributed to the 
increase in conversion of $SO$ to $SO_2$ via radiative association with atomic oxygen and then freeze out of $SO_2$ onto dust grains \citep{Pacheco2016}.
In HD 100546, the density and temperature of the disk may have been perturbed 
creating the conditions for the localised formation of the observed asymmetric SO. However, chemical modeling of warped disks
and associated temperature perturbations is required to confirm this hypothesis.

From observations of cometary volatiles in the particular, for the case of 67P, the total abundance of
sulphur-bearing species detected is consistent with the 
solar abundance of sulphur \citep{2016MNRAS.462S.253C}.
This means that if our solar system is typical then the observed depletion of sulphur in circumstellar regions  
may be an observational effect as we have not been able to detect the various forms of sulphur.
The form of the sulphur, whether it resides in refractory or volatile form in planet-forming disks is still an open question.
For the SO we detect in HD 100546 it is unclear as to its origin, e.g., if it is a result of the volatile reactions described above
or whether it has been released from refractory materials due to a shock as suggested by the abundance we determine. 
Further observations may make this clearer.

\subsection{What is the SO tracing?}

The influence of the massive companion at approximately 10~au in the disk may cause the disk velocity structure to depart from 
simple Keplerian rotation in the inner region. The protoplanet embedded in the disk at 
50~au may also have an effect \citep{2006ApJ...640.1078Q}.
If the disk is warped, the line of sight inclination will vary radially changing the velocity structure.
Previous observations with APEX show the $^{12}CO$ emission from the HD 100546 protoplanetary disk is 
asymmetric \citep{Panic2010} suggesting that one side of the outer disk is colder by 10-20 K than the other or
that there is a shadow on the outer disk caused by a warped geometry of the inner disk.
Shadows resulting from disk warps and their effect on gas kinematics have been observed in a few other sources  
e.g. HD~142527 \citep{Casassus:2015hm}.
There is evidence for a possible warp in the inner 100~au of the HD~100546 disk from 
a detailed study of the CO $J=3-2$ kinematics \citep{walsh2017}.
The spatial resolution of the HD~100546 Cycle 0 observations is limited to approximately 100~au along the minor axis of the disk
so it is unclear what is causing the non-Keplerian motions. Since previous observations point towards 
this star hosting at least one massive companion, a warped disk is a favoured hypothesis. 
A warp would directly expose the north-east side of the disk
to heating by the central star, creating locally the conditions for the formation of SO and the launching of a disk wind. 
The non-Keplerian motions in the inner 100~au of this disk could explain the discrepancy in the kinematics of the 
best fit model for the SO and the observations.  
   
Disk dispersal is predicted to occur on a timescale up to 10 times shorter than observed disk lifetimes \citep{Alexander2014}. This process
limits the time available for giant plant formation, decreases the gas to dust
ratio in the disk, and the mass loss will have an effect on the chemical content of the disk.
In planet-forming Class II disks, jets/outflows are not the main driver of disk dispersal. Instead
the removal of angular momentum from the disk material can be achieved by slower disk winds ($<$~30~km~s$^{-1}$).
Photoevaporative disk winds are thought to be the primary disk dispersal mechanism \citep{Alexander2014}.
MHD disk winds also drive disk dispersal 
but are less well understood \citep{2017arXiv170400214E}.  
Evidence of photoevaporative disk winds has been 
detected from a number of sources in the form of blue-shifted (up to 10~km~s$^{-1}$) line profiles of forbidden line emission in the optical
\citep[e.g.][]{Pascucci2011, 2016MNRAS.460.3472E}. In addition to this, ALMA 
observations show a spatially resolved molecular disk wind originating from the HD 163296 disk \citep{Klaassen2013},
and a molecular protostellar outflow from TMC1A launched by a disk wind originating from a Keplerian disk \citep{2016Natur.540..406B}.
We checked for any large scale or high velocity ($>$10~km~s$^{-1}$) CO or SO emission from HD100546 but none was detected. 
Gas launched from the HD 100546 disk surface with a velocity of a 
few km~s$^{-1}$ would have a blue shift along the line of sight to the observer and could account for the blue-shifted emission 
we observe. The mechanism for launching this material and why it is traced in the SO emission is unclear.
The lack of observed excess blue-shifted CO emission is due to the SO originating from a layer in the disk that is higher than the emitting layer of the CO $J=3-2$ (E$_u=$33.19~K) gas.
To see this effect in CO will require observations of higher J lines which will be tracing the warmer gas in the atmosphere.
We surmise that the red-shifted counterpart of the disk wind, launching from the far side of the disk, is obscured by the optically thick dust disk.
Determination of the chemical origin of the SO will help to shed light on whether the wind is MHD driven 
(ion-molecule chemistry) or photoevaporative (photon-dominated chemistry) in nature.

An alternative explanation for the SO emission is that it is the result of an accretion shock 
due to a circumplanetary disk. The position angle of the SO emission coincides with the
observed infrared point source in the disk at approximately 50~au that has been attributed to a protoplanet \citep{Quanz2013,Currie2015}.

\subsection{Conclusions}

We have shown that SO is detectable in protoplanetary disks with ALMA uncovering a sulphur reservoir in the HD 100546 protoplanetary disk.
In addition, we have shown that SO may be a tracer of a molecular disk wind.
New data with better spatial and spectral resolution are required to 
truly disentangle the disk and wind components of the emission.
This will allow for an accurate determination of the spatial distribution of the disk emission and the launch region of the wind. 
Observations of sulphur-bearing species in this disk and others will help to constrain the 
fraction of the cosmic abundance of sulphur partitioned in the refractory and the volatile
materials. Transition disks are important targets as their cavities may expose the hidden sulphur. 

\begin{acknowledgements}
This paper makes use of the following ALMA data: ADS/JAO.ALMA\#2011.0.00863.S ALMA is a partnership of
ESO (representing its member states), NSF (USA) and NINS (Japan), together with NRC (Canada), NSC and ASIAA 
(Taiwan), and KASI (Republic of Korea), in cooperation with the Republic of Chile. The Joint ALMA Observatory
is operated by ESO, AUI/NRAO and NAOJ. 
A.B. acknowledges the studentship funded by the Science and Technology Facilities Council of the United Kingdom (STFC).
C.W. acknowledges financial support from the Netherlands Organisation for Scientific Research (NWO, grant 639.041.335) and start-up funds from the University of Leeds, UK.
This work is partly funded by STFC consolidated grant number \#\#\#.
R.A.L. gratefully acknowledges funding from the NRAO Student Observing Support program.
A.J. is supported by the DISCSIM project, grant agreement 341137 funded by the European Research Council under ERC-2013-ADG.

\end{acknowledgements}

\bibliography{SO_HD100546_final.bib}

\begin{appendix}
\onecolumn

\section{Individual line profiles}
\begin{figure}[h!]
\includegraphics[width=0.49\hsize]{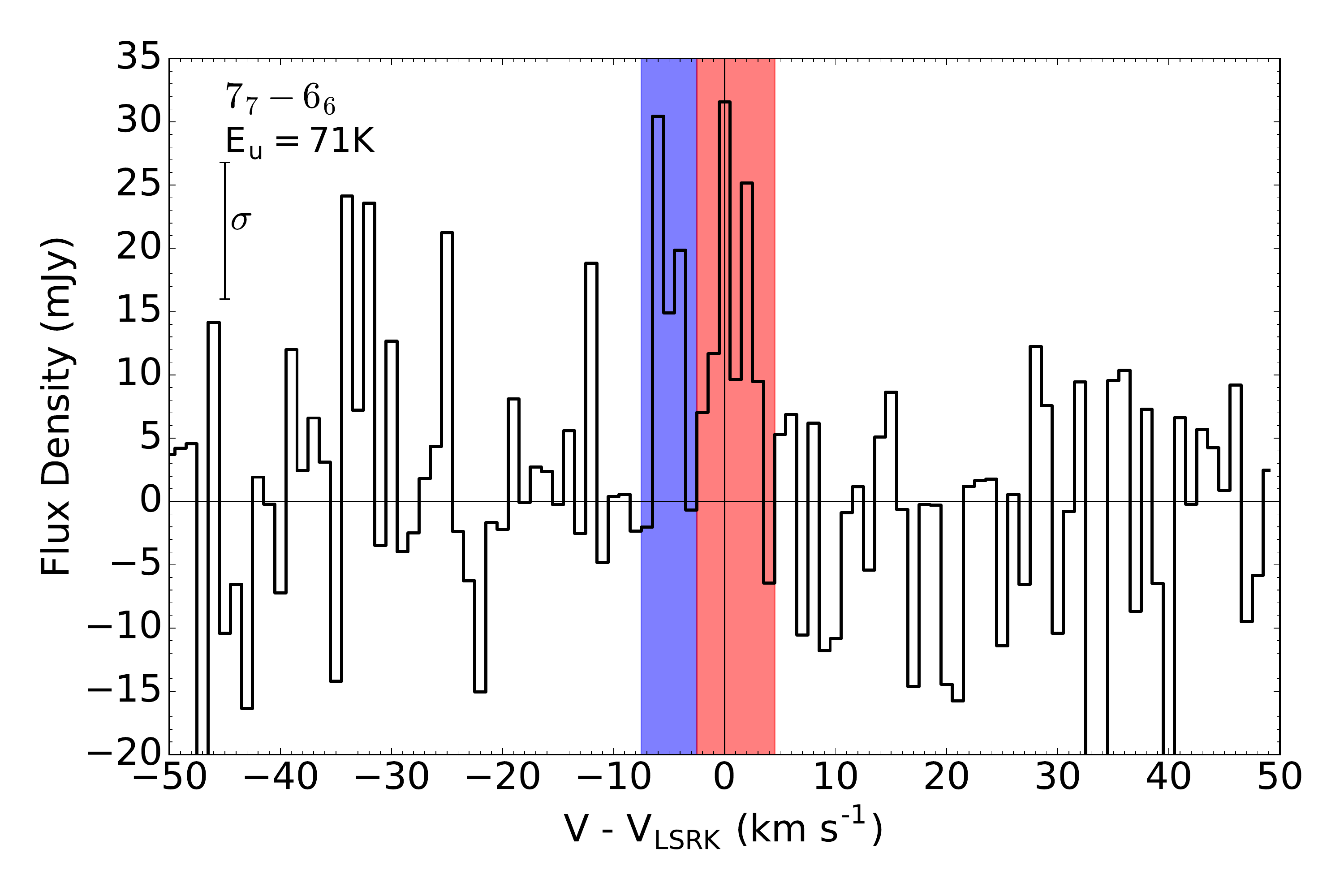}
\includegraphics[width=0.49\hsize]{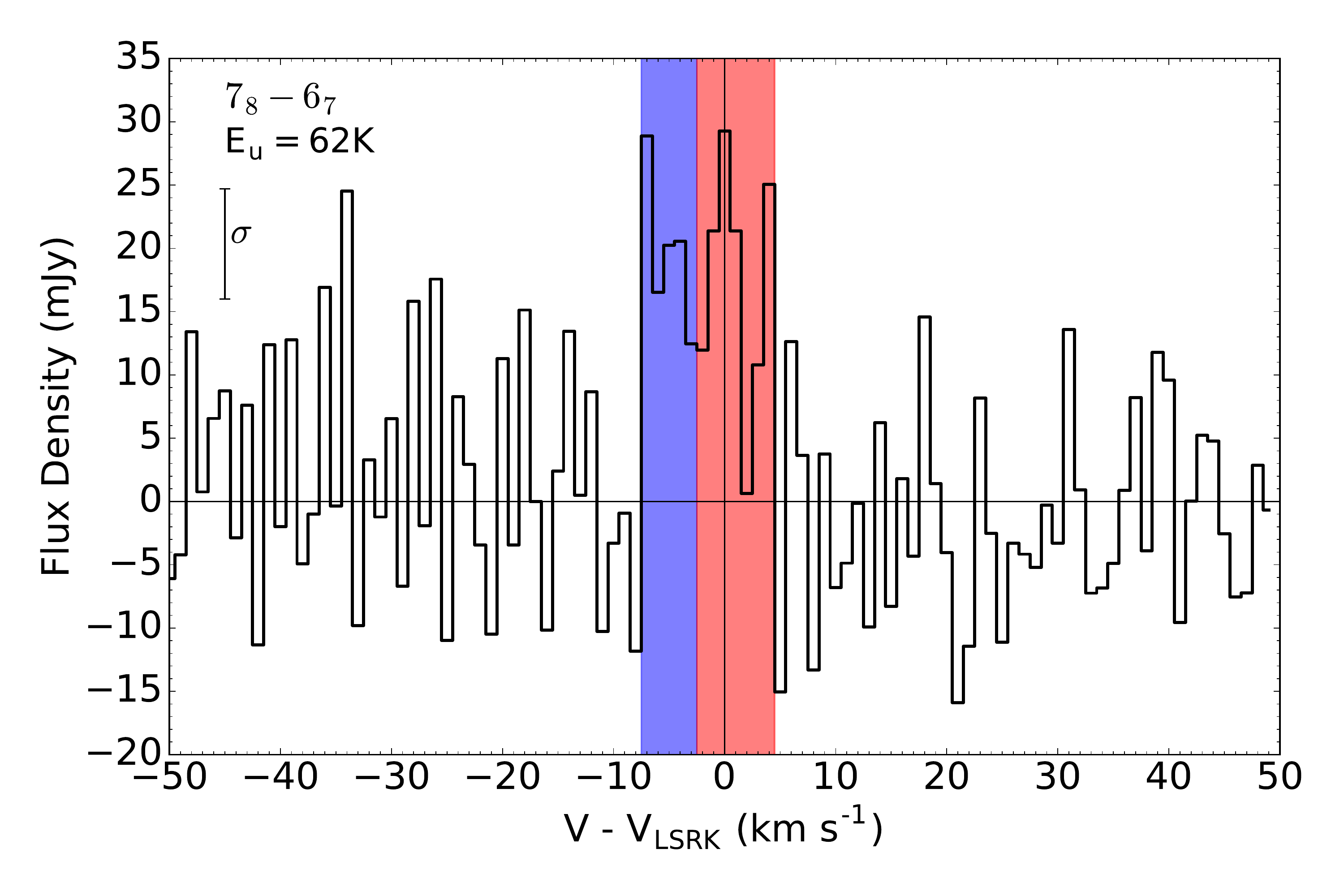}
\caption{Line profiles of the individual J=$7_{7}-6_{6}$ (left) and J=$7_{8}-6_{7}$ (right) transitions extracted from within the 3$\sigma$ extent of their respective intensity maps.
Both line profiles reach a S/N of 3 with r.m.s. noise of 10.8~mJy and 8.7~mJy respectively.
Highlighted in red and blue are the velocity ranges of emission used in the moment maps in Figure 4.}
\end{figure}

\section{SO and CO J=3-2 line profiles}
\begin{figure}[h!]
  \centering
  \includegraphics[width=0.5\hsize]{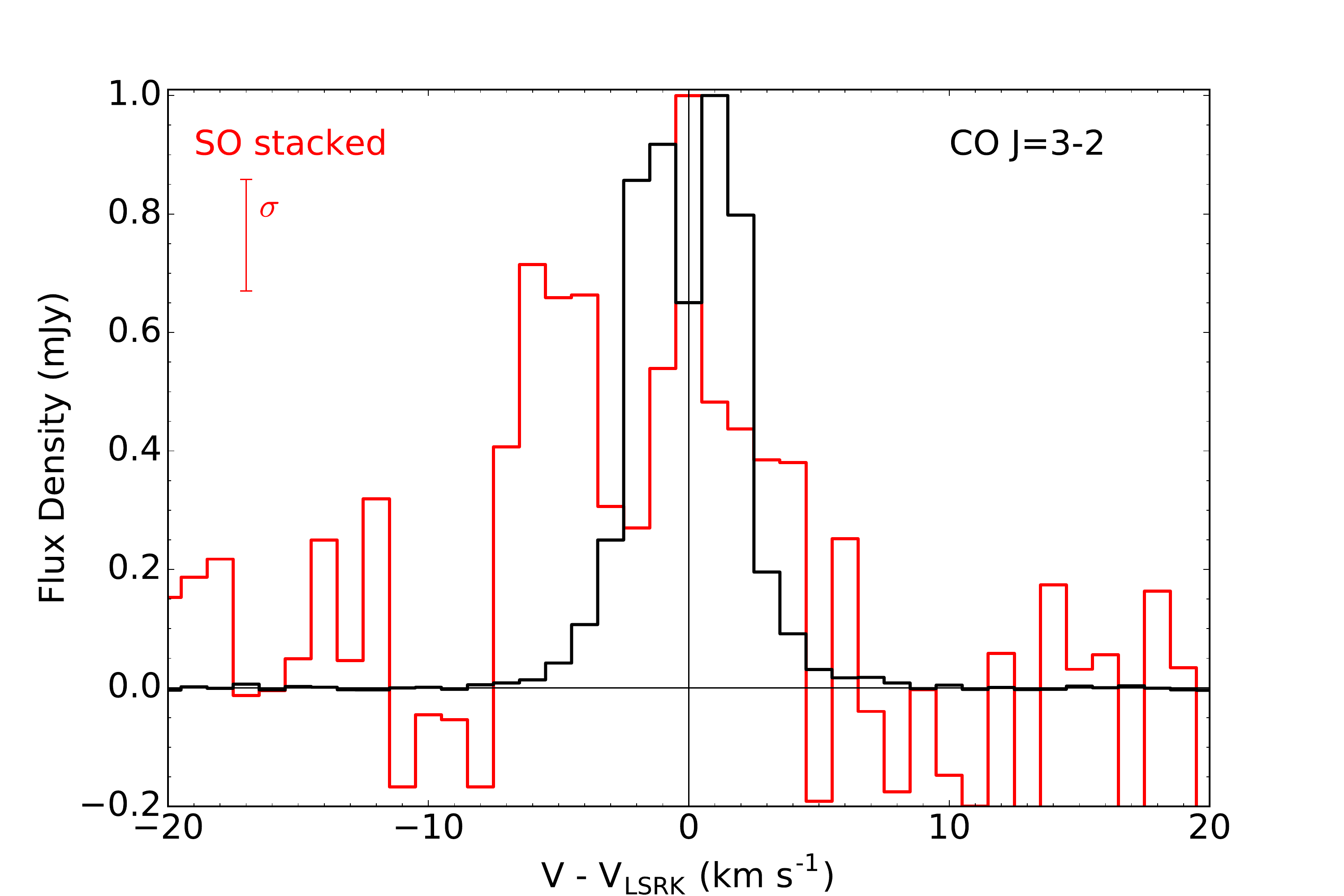}
  \caption{SO stacked (red) line profile and CO $J=3-2$ (black) line profile extracted from within the 3$\sigma$ extent of their respective stacked integrated intensity maps.
  The SO stacked reaches a S/N of 5 and the CO $J=3-2$ reaches a S/N of 375. Note that the r.m.s. for the CO $J=3-2$ is not visible on this scale (81~mJy).}
  \label{}
\end{figure}

\newpage
\section{Stacked moment maps}

\begin{figure*}[h!]
\includegraphics[width=0.5\hsize]{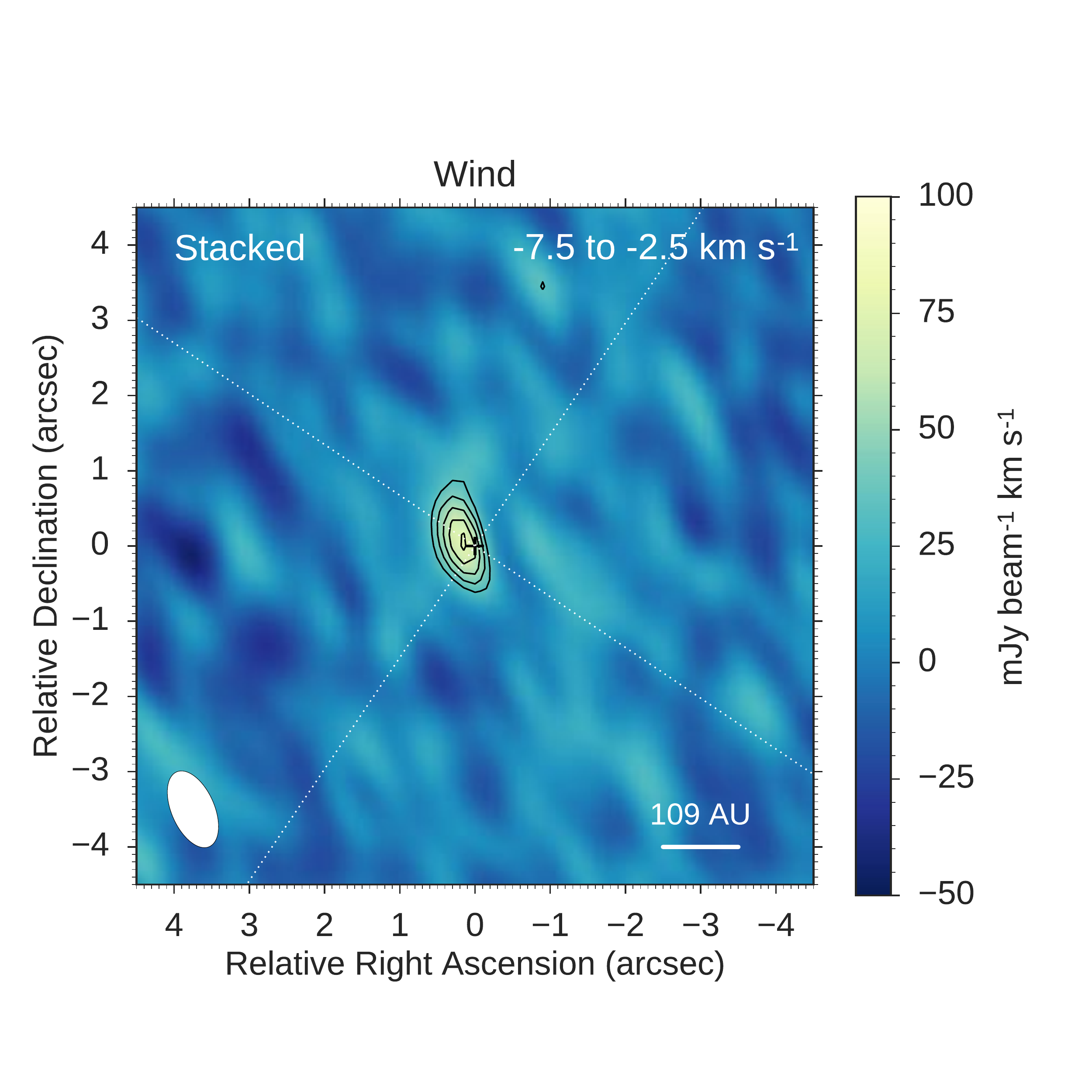}
\includegraphics[width=0.5\hsize]{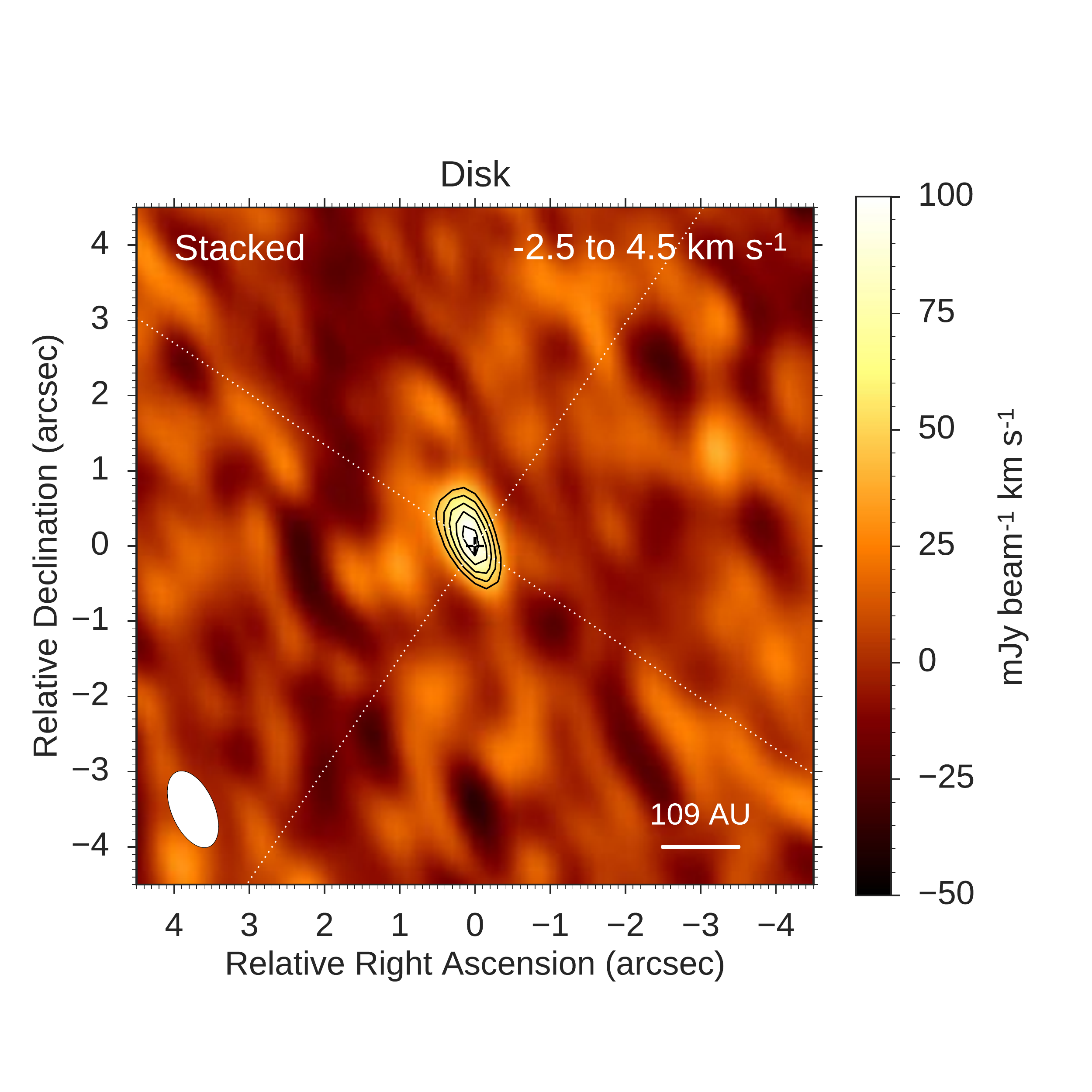}
\caption{Integrated intensity maps of the stacked SO emission over two velocity ranges:
-7.5~km~s$^{-1}$to -2.5~km~s$^{-1}$ (left) with a S/N of 7 and -2.5~km~s$^{-1}$to 4.5~km~s$^{-1}$ (right) with a S/N of 9.
The r.m.s. noise these maps is 11 and 14~mJy~beam$^{-1}$~km~s$^{-1}$ and the peak emission is 80 and 123~mJy~beam$^{-1}$~km~s$^{-1}$ respectively.
The black contours are at intervals of $\sigma$ going from 3$\sigma$ to peak.}
\end{figure*}

\newpage
\section{RADEX Line Ratios}
\begin{table}[h!]
\centering
\caption{Model and observed line ratios for detected SO transitions.}
\begin{tabular}{ccccc}
\hline\hline
\multicolumn{5}{c}{Observed line ratios}\\
\hline
- & -    & 1.5$\pm$0.4 &  1.6$\pm$0.4   & 1.1$\pm$0.4  \\
\hline\hline
\multicolumn{5}{c}{Radex model line ratios}\\
\hline 
n$_{H_2}$ (cm$^{-3}$)    & T$_{k}$ (K) & J=$7_{8}-6_{7}$/J=$7_{7}-6_{6}$ & J=$7_{8}-6_{7}$/J=$8_{7}-7_{6}$ & J=$7_{7}-6_{6}$/J=$8_{7}-7_{6}$ \\
\hline
&&&&\\
$10^5$    & 25     & 1.909   & 7.000   & 3.667   \\
$10^6$    & 25     & 1.726   & 3.982   & 2.296   \\
$10^7$    & 25     & 1.607   & 2.481   & 1.544   \\
$10^8$    & 25     & 1.587   & 2.302   & 1.451   \\
$10^9$    & 25     & 1.583   & 2.288   & 1.446   \\
$10^{10}$ & 25     & 1.583   & 2.288   & 1.446   \\
&&&&\\
$10^5$    & 50     & 1.659   & 4.056   & 2.444   \\
$10^6$    & 50     & 1.489   & 2.285   & 1.535   \\
$10^7$    & 50     & 1.386   & 1.530   & 1.104   \\
$10^8$    & 50     & 1.367   & 1.449   & 1.060   \\
$10^9$    & 50     & 1.364   & 1.442   & 1.058   \\
$10^{10}$ & 50     & 1.364   & 1.442   & 1.058   \\
&&&&\\
$10^5$    & 100    & 1.546   & 2.930   & 1.895   \\
$10^6$    & 100    & 1.400   & 1.683   & 1.203   \\
$10^7$    & 100    & 1.296   & 1.197   & 0.924   \\
$10^8$    & 100    & 1.273   & 1.150   & 0.903   \\
$10^9$    & 100    & 1.269   & 1.144   & 0.902   \\
$10^{10}$ & 100    & 1.270   & 1.145   & 0.902   \\
&&&&\\
$10^5$    & 250    & 1.555   & 2.348   & 1.510   \\
$10^6$    & 250    & 1.371   & 1.318   & 0.961   \\
$10^7$    & 250    & 1.239   & 1.015   & 0.819   \\
$10^8$    & 250    & 1.212   & 0.990   & 0.816   \\
$10^9$    & 250    & 1.211   & 0.984   & 0.812   \\
$10^{10}$ & 250    & 1.203   & 0.978   & 0.813   \\
&&&&\\
$10^5$    & 500    & 1.551   & 2.182   & 1.407   \\
$10^6$    & 500    & 1.358   & 1.203   & 0.886   \\
$10^7$    & 500    & 1.218   & 0.962   & 0.790   \\
$10^8$    & 500    & 1.192   & 0.939   & 0.788   \\
$10^9$    & 500    & 1.191   & 0.939   & 0.789   \\
$10^{10}$ & 500    & 1.193   & 0.939   & 0.787   \\
&&&&\\
\hline 
\end{tabular}
\end{table}

\newpage
\section{Individual line profiles with LIME disk model line profiles}
\begin{figure}[h!]
\includegraphics[width=0.49\hsize]{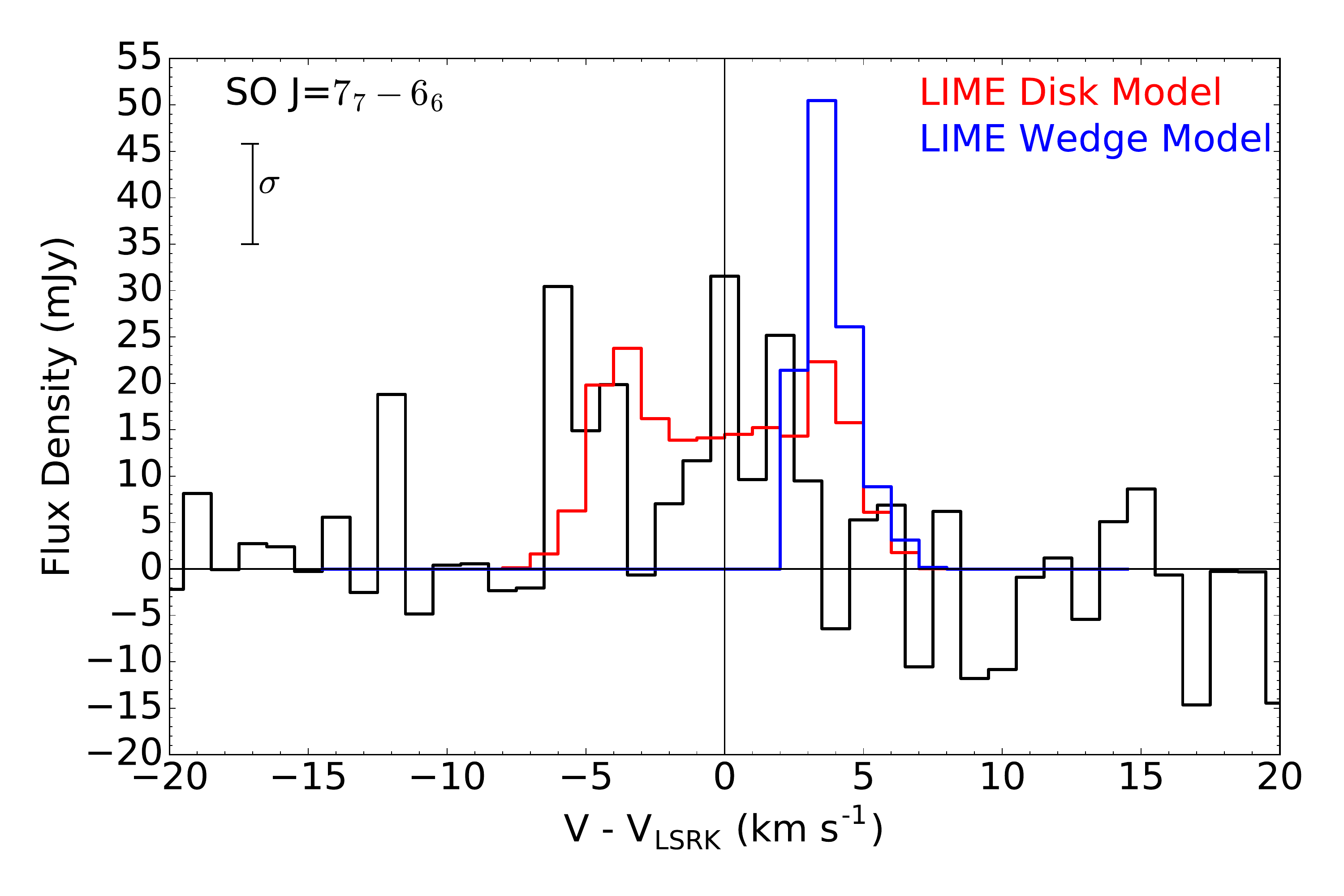}
\includegraphics[width=0.49\hsize]{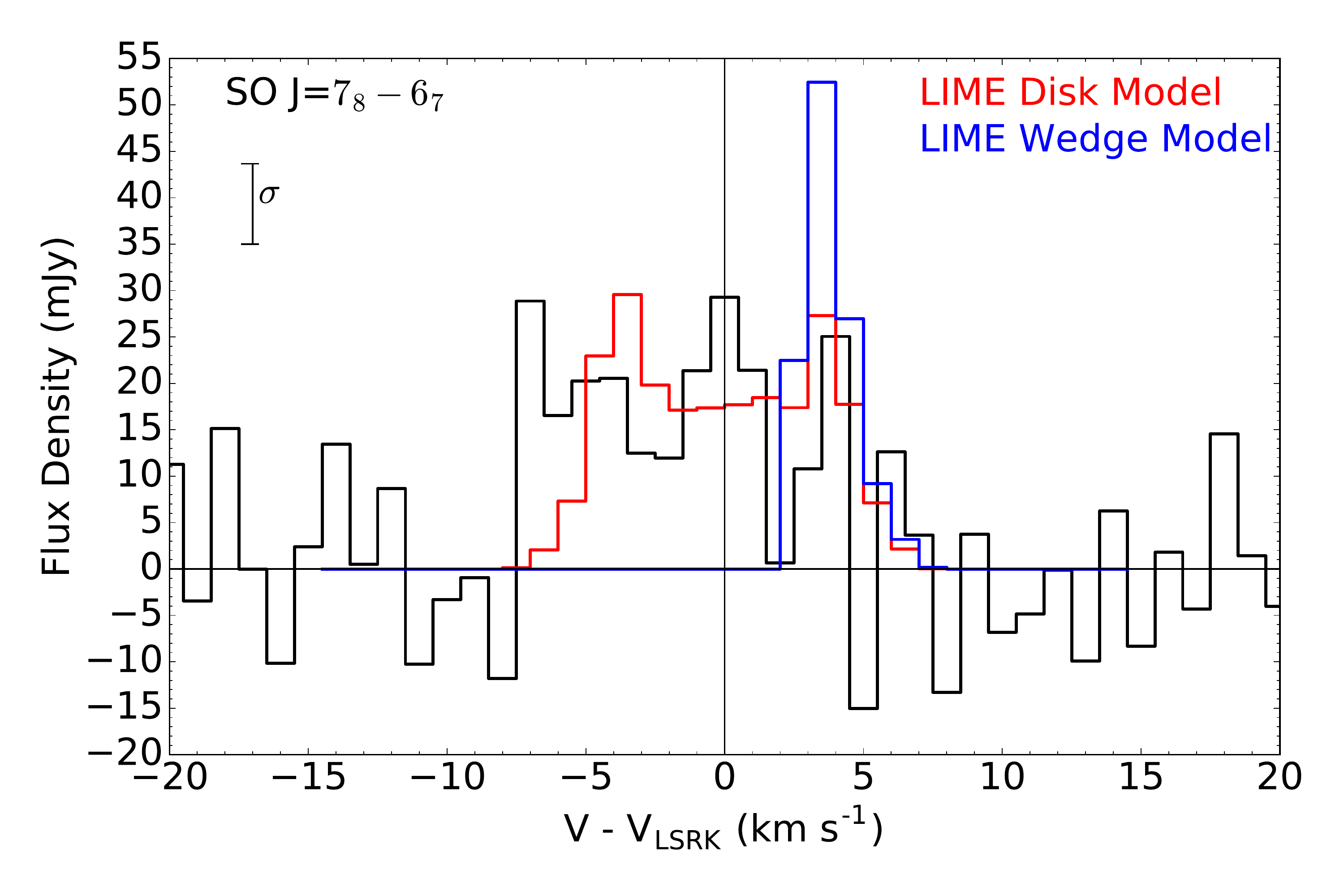}
\caption{Line profiles of the individual J=$7_{7}-6_{6}$ (left) and J=$7_{8}-6_{7}$ (right) transitions with the LIME model line profiles
plotted on the top: disk model (red) and wedge model (blue).}
\end{figure}

\end{appendix}
\end{document}